\documentclass[11 pt,a4 paper]{article}
\usepackage{jheppub}
\pdfoutput=1

\usepackage{amsmath}
\usepackage{amssymb}
\usepackage{amsfonts}
\usepackage{braket}
\usepackage[utf8x]{inputenc}

\DeclareMathOperator{\Tr}{Tr}

\title{Eigenstate Thermalization Hypothesis and Approximate Quantum Error Correction}
\author[a,b]{Ning Bao}
\author[a]{and Newton Cheng}

\affiliation[a]{Berkeley Center for Theoretical Physics, Berkeley, CA, 94720, USA}
\affiliation[b]{Computational Science Initiative, Brookhaven National Lab, Upton, NY, 11973, USA}

\emailAdd{ningbao75@gmail.com}
\emailAdd{newtoncheng@berkeley.edu}

\abstract{The eigenstate thermalization hypothesis (ETH) is a powerful conjecture for understanding how statistical mechanics emerges in a large class of many-body quantum systems. It has also been interpreted in a CFT context, and, in particular, holographic CFTs are expected to satisfy ETH. Recently, it was observed that the ETH condition corresponds to a necessary and sufficient condition for an approximate quantum error correcting code (AQECC), implying the presence of AQECCs in systems satisfying ETH. In this paper, we explore the properties of ETH as an error correcting code and show that there exists an explicit universal recovery channel for the code. Based on the analysis, we discuss a generalization that all chaotic theories contain error correcting codes. We then specialize to AdS/CFT to demonstrate the possibility of total bulk reconstruction in black holes with a well-defined macroscopic geometry. When combined with the existing AdS/CFT error correction story, this shows that black holes are enormously robust against erasure errors.}

\begin{document}
\maketitle
\section{Introduction}
The AdS/CFT correspondence states that a theory of quantum gravity in $(d+1)$-dimensional anti-de Sitter space is dual to a $d$-dimensional conformal field theory that ``lives'' on the boundary of the space. The theories are dual in the sense  we can write down a ``dictionary'' that allows us to map observables in one theory to the other, while still preserving algebraic relations, i.e. describing the same physics. It was suggested in \cite{Almheiri:2014lwa} that the mechanism by which this holographic duality occurs is best understood in the framework of quantum error correcting codes (QECCs). In this perspective, the dictionary is really the map that encodes the ``logical'' bulk degrees of freedom into the ``physical'' CFT system. These ``holographic codes'' are robust against erasure errors on the boundary, and it has been shown that the properties of these codes can be used to explain various strange properties and paradoxes of the correspondence \cite{Harlow:2016vwg}.

However, the connection to QECCs has yielded more fruit than the original conjecture in \cite{Almheiri:2014lwa} -- recent work has applied the tools of error correction to study topics such as the link between entanglement and spacetime \cite{Pastawski:2015qua, Akers:2018fow,Bao:2018pvs} and black holes \cite{Verlinde:2012cy,Almheiri:2018xdw}. In this work, we continue in this direction by studying the error correcting properties of CFTs in the AdS/CFT correspondence. More generally, we analyze the observation made in \cite{Brandao:2017irx} that systems satisfying the eigenstate thermalization hypothesis (ETH) contain approximate error correcting codes in their spectrums. In doing so, we find a connection between generic chaotic theories, of which holographic CFTs are an example, and quantum error correction. This connection then has deep implications for black holes in AdS/CFT.

This is not a new idea, and has permeated through the literature in different forms. The essential idea behind treating chaos, thought of as ``randomness,'' as a kind of error correcting code has been understood in the context of information channels \cite{Hayden:2008a,Hayden:2008b,Horodecki:2008}, where it was shown that random subspaces can provide an encoding that achieves the capacity of a noisy channel. A similar idea was used in \cite{Hosur:2015ylk} to show that the input state to a system that scrambles quantum information, defined by the decay of out-of-time-order correlators, can be recovered due to the random nature of the time evolution. In the classic paper from Hayden and Preskill \cite{Hayden:2007cs}, the equivalence of scrambling in random unitary circuits and protection from erasure is the key took in showing that a black hole acts as a ``mirror'' that very quickly radiates quantum information of objects thrown into it after radiating half its mass. Recent work \cite{Yoshida:2018ybz,Yoshida:2019qqw} has further explored scrambling in the context of the black hole information paradox and firewalls. In this work, we extend the connection between QEC and scrambling to chaotic theories in general, as well as sharpening the connection by taking an explicit approach from a coding perspective applied to ETH. We then apply the ideas to study black holes in AdS/CFT.

The outline of the paper is as follows: in section \ref{prelim}, we review the eigenstate thermalization hypothesis, its interpretation in the CFT context, as well as generic approximate quantum error correction. In section \ref{ethasaqecc}, we interpret and analyze ETH as an approximate error correcting code. We show that the theory of universal recovery channels can be applied to the ETH code, and discuss the proposition that all chaotic theories necessarily contain error correcting codes. In section \ref{bhandtbr}, we apply the ETH code to analyze black holes in AdS/CFT and demonstrate that total bulk reconstruction is possible. We discuss how the properties of the ETH code are very different, but complementary, to the codes in the usual AdS/CFT story, and affords black holes an enormous amount of protecting against erasures. In section \ref{disc}, we discuss connections to recent work as well as possible future directions.

To make notation simpler, we always take the base of a logarithm to be the dimension of the smallest subsystem $\mathcal{H}_s$ of the full system $\mathcal{H}$, e.g. base 2 when taking about a system composed of qubits. When we want to be specific, we will use $s = \dim \mathcal{H}_s$.

\section{Preliminaries}\label{prelim}
\subsection{Eigenstate thermalization hypothesis}
The eigenstate thermalization hypothesis is formulated as an ansatz for the matrix elements of operators in the energy eigenbasis of a non-integrable or ``chaotic'' Hamiltonian:
\begin{equation}\label{eth}
	\braket{E_m|\mathcal{O}|E_n} = O_{\mathcal{O}}(\bar{E})\delta_{mn} + e^{-S(\bar{E})/2}f_\mathcal{O}(\bar{E},\omega)R_{mn},
\end{equation}
where $\bar{E} = \frac{E_m+E_n}{2}$ and $\omega = \frac{E_n-E_m}{2}$. In this expression, $S(\bar{E})$ is the thermodynamic entropy at energy $\bar{E}$, the functions $O_{\mathcal{O}}(\bar{E})$,  $f_\mathcal{O}(\bar{E},\omega)$ are smooth and depend on the operator $\mathcal{O}$, and $R_{mn}$ is a random distribution with zero mean and unit variance. In the original statement, the operator $\mathcal{O}$ in question should be a ``few-body'' local observable, in the sense that $\mathcal{O}$ is a local $n$-body observable in a system of $N$ particles with $N \gg n$ -- these are the observables that can be studied in the lab. For example, it is clear that the projection operator $P = \ket{m}\bra{m}$ does not satisfy the ETH ansatz. The power of this statement comes when we select a small energy band in the bulk of the spectrum $[E-\Delta E/2, E + \Delta E/2]$. The width $\Delta E$ is selected to be much larger than the energy spacing, which is exponentially suppressed in the system size $N$:
\begin{equation}
    |E_{i+1} - E_{i}| \propto e^{-cN}\ \text{for all $\{E_i\}$ in the band},
\end{equation}
for some constant $c$. Then if $O_{\mathcal{O}}(\bar{E})$ takes on the value of the microcanonical average of $O$ at energy $\bar{E}$ and assuming the entropy is extensive, we get
\begin{equation}\label{eth2}
	\Tr_{A}(\rho_A\mathcal{O}) = \Tr_{A}(\rho_{A,\text{therm}}\mathcal{O})
\end{equation}
in the thermodynamic limit $V \to \infty$. The thermal state is reduced on the canonical ensemble
\begin{equation}
	\rho_{A,\text{therm}} = \frac{1}{Z}\Tr_{\bar{A}}\left(e^{-\beta H}\right)
\end{equation}
The qualitative description of ETH is that we imagine splitting our system into two subsystems, one smaller than the other. Observables in the smaller subregion are thermalized by the larger subregion, which acts like a heat bath at some effective temperature. We refer the reader to \cite{DAlessio:2016rwt} and \cite{Deutsch:2018a} for a more detailed discussion of the context in which ETH is formulated. 

Although the original formulation of ETH only applied to few-body local observables, we will assume the stance taken in \cite{Garrison:2015lva} as to which operators satisfy ETH. Imagine that we have split the system into subregions $A,\bar{A}$ with $V_{\bar{A}} > V_A$, and that the reduced density matrix on $A$ is thermal at a temperature $T = 1/\beta$. Now divide operators into two classes based on the behavior of their expectation values in the limit $V_A \to \infty$: Class I operators have expectation values that include contributions only from eigenstates of $H_A$ with energy densities corresponding to $\beta$, while Class II operators are all operators that are not Class I. All local operators are certainly Class I, so their formulation encapsulates the original ETH. They conjecture that in the thermodynamic limit $V_A/V \to 0$ as $V_A,V \to \infty$, ETH should hold for all operators, irrespective of class. For our purposes, we are interested in the thermodynamic limit where $V_A/V \to f < 1/2$ as $V_A,V \to \infty$. In this case, they conjecture that almost all Class I operators, and all Class II operators below a critical energy density satisfy ETH. Operators that involve energy conservation, e.g. the variance of the energy, do not satisfy ETH, because they are proportional to an extra factor of $(1-f)$. More precisely, any operator in the set spanned by $\{H_A,H_A^2,\ldots,H_A^n\}$, where $n$ is the size of the Hilbert space, is not expected to satisfy ETH.

In \cite{Dymarsky:2016ntg,Lashkari:2017hwq}, the ETH condition was transplanted into the CFT context. Consider a $d$-dimensional CFT living on a $(d-1)$-dimensional hypersphere $S^{d-1}$ of radius $L$.   We consider the reduced density matrix of a CFT energy eigenstate on a subsystem $A$ smaller than its complement $\rho_A^i = \Tr_{\bar{A}}\ket{E_i}\bra{E_i}$. Assuming that one satisfies the original ETH condition:
\begin{enumerate}
	\item There exists a universal density matrix (``ETH density matrix'') $\rho_A(E_i)$ that varies smoothly with $E_i$ such that
	\begin{equation}\label{subeth}
		\frac{1}{2}||\rho_A^i - \rho_{\text{ETH}}(A,E_i)||_1 \sim O(e^{-O(S(E_i))}),
	\end{equation}
	where $\frac{1}{2}||\rho - \sigma||_1$ is the trace distance $\frac{1}{2}\Tr|\rho-\sigma|$.
	\item The off-diagonal elements $\rho_A^{ij} = \Tr_{\bar{A}}\ket{E_i}\bra{E_j}$ of the reduced density matrix satisfy
	\begin{equation}
		||\rho_A^{ij}||_1 \sim O(e^{-O(S(\bar{E}))}),
	\end{equation}
	where $\bar{E} = \frac{E_i+E_j}{2}$.
\end{enumerate}
This formulation of ETH says that the matrix elements of the state on $A$ are exponentially close to a universal ETH density matrix, and the off-diagonal elements are exponentially suppressed. Assuming (\ref{eth}), one can then show that expectation values of local operators match onto thermal expectation values. Based on this equivalence, the authors of \cite{Lashkari:2017hwq} argue that the ETH density matrix is well-approximated by the reduced density matrix of a global thermal state in the canonical ensemble:
\begin{equation}\label{ethstate}
	\rho_{\text{ETH}} = \frac{1}{Z}\Tr_{\bar{A}}\left(e^{-\beta H}\right) + O(1/\sqrt{L}),
\end{equation}
which formalizes the qualitative description that reduced states in chaotic systems look thermal in the thermodynamic limit. In the thermodynamic limit $L \to \infty$, we get precisely the same result as (\ref{eth2}). We will assume that the result in \cite{Garrison:2015lva} that for $f < 1/2$, all local operators not involving energy conservation obey ETH also holds for chaotic CFTs.

Let us be careful about what ETH says: ETH does not say that the reduced density matrix of a subregion is exponentially close in trace-distance to a reduced thermal state. Rather, the state is exponentially close to some universal density matrix. Therefore, when we calculate any expectation values in the energy eigenstates of a band of energies with exponentially small spacing, the expectation value under any of the eigenstates in the band is well-approximated by any other eigenstate in the band to exponential accuracy. However, the reduced state is close to a true thermal state only to polynomial accuracy; this has to do with the fact that the equivalence of thermodynamic ensembles in the thermodynamic limit occurs with polynomial corrections in the system size. When we take the thermodynamic limit, both statements become exact, as expected.

\subsection{Approximate quantum error correction}
We now provide a brief introduction and review of aspects of generic (approximate) quantum error correction. A quantum error correcting code is defined as a subsystem $\mathcal{C} \subseteq \mathcal{H}$ of the full Hilbert space, called a ``code subspace'' (the code subspace is not always a true subspace). The standard terminology is that one encodes the ``logical'' code subspace into a larger ``physical'' Hilbert space, which then protects the logical system from errors. Let $\mathcal{S}(\mathcal{H})$ denote the space of states $\rho$ on a Hilbert space $\mathcal{H}$. A quantum noise channel $\mathcal{N}:\mathcal{S}(\mathcal{C})\to\mathcal{S}(\mathcal{H})$ can be defined by a set of operator elements, also called \emph{Kraus operators}, $\{E_i\}$ that satisfy
\begin{equation}
	\sum_i E_i^\dagger E_i = I.
\end{equation}
We restrict our attention to channels that are completely positive and trace-preserving. The effect of the channel on a state in the code subspace $\rho \in \mathcal{S}(\mathcal{C})$ can then be represented as an operator sum:
\begin{equation}
	\mathcal{N}(\rho) = \sum_i E_i\rho E_i^\dagger.
\end{equation}
The error is said to be correctable if there exists a recovery channel $\mathcal{R}:\mathcal{S}(\mathcal{H})\to\mathcal{S}(\mathcal{C})$ that satisfies:
\begin{equation}
	(\mathcal{R}\circ \mathcal{N})(\rho) = \rho,\qquad \text{for all }\rho \in \mathcal{S}(\mathcal{C}).
\end{equation}
That is, the recovery channel reverses the effect of the noise channel, correcting the error induced by the noise. Depending on the properties of the code, it may correct only particular noise channels or arbitrary noise channels. The properties of a code for a given channel are neatly summarized by the notation $[[n,k,d]]$, where the ``length'' $n = \log \dim \mathcal{H}$ is the size of the physical system, $k = \log \dim \mathcal{C}$ is the size of the code subspace, and the distance $d$ is the smallest region that cannot be corrected. In a qubit system, $n$ is the total number of qubits, $k$ is the number of encoded qubits, and the code can correct errors on at most $d-1$ qubits. Good codes will have the largest possible $k$ and $d$ for a given $n$. 

The noise channel that we will be most interested in is the erasure channel, which we intuitively imagine as erasing some portion of the physical Hilbert space and replacing it with some arbitrary state. More precisely, we consider a Hilbert space that admits a bipartite tensor factorization $\mathcal{H} = \mathcal{H}_A \otimes \mathcal{H}_{\bar{A}}$, and the noise channel erasing ${A}$ has the representation:
\begin{equation}\label{erasure}
	\mathcal{N}(\rho) = \Tr_{{A}}(\rho) \otimes \frac{1}{\dim \mathcal{H}_{{A}}}I_{{A}}.
\end{equation}
We have traced over the ${A}$ subsystem, and replaced it with the maximally mixed state that carries no quantum information.

The question of what errors can be corrected for a given $\mathcal{C}$ was answered in \cite{Knill:1996ny}, who gave the necessary and sufficient Knill-Laflamme condition for error correction of a channel with operator elements $\{E_i\}$:
\begin{equation}\label{knill}
PE_i^\dagger E_j P = C_{ij}P,
\end{equation}
where $P$ is the projector onto the code subspace and $C_{ij}$ is a hermitian matrix. It is useful to choose an orthonormal basis $\{\ket{\psi_i}\}$ for $\mathcal{C}$ so that the condition above becomes
\begin{equation}\label{knill2}
\braket{\psi_m|E_i^\dagger E_j|\psi_n} = C_{ij}\delta_{mn}
\end{equation}
The intuition behind error correction is that the procedure of encoding the logical data into a physical space creates redundancy. When a correctable error occurs, the error has only affected the redundant information, so it is possible to recover the logical information. 

The conditions (\ref{knill}), (\ref{knill2}) are for exact error correction. Comparing (\ref{knill2}) and (\ref{eth}), they differ by the exponentially-suppressed fluctuation term. However, there is no guarantee that the exact correction conditions are robust in the sense that approximate satisfaction implies approximate correction. This turns out to be true, and was worked out in \cite{Beny:2010a}. Before presenting their result, we briefly discuss how to compare states to formalize the notion of ``approximately'' recovering a state. The simplest distance measure between states is the trace distance:
\begin{equation}
	\frac{1}{2}||\rho-\sigma||_1 = \frac{1}{2}\Tr|\rho-\sigma|.
\end{equation}
This quantity takes values on the interval $[0,1]$, is 1 if and only if $\rho$ and $\sigma$ are supported on orthogonal subspaces, and is 0 if and only if $\rho = \sigma$. Clearly we want this quantity to be as small as possible. Another common way to compare states is the state fidelity:
\begin{equation}
	f(\rho,\sigma) = \Tr\left[\sqrt{\sqrt{\sigma}\rho\sqrt{\sigma}}\right]^2.
\end{equation}
This quantity takes values $f \in [0,1]$, equals 1 if and only if $\rho = \sigma$, and equals 0 if and only if $\rho$ and $\sigma$ are orthogonal. These two quantities are related by the Fuchs-van de Graaf inequality:
\begin{equation}
	1-f(\rho,\sigma) \leq \frac{1}{2}||\rho-\sigma||_1 \leq \sqrt{1-f^2(\rho,\sigma)}.
\end{equation}
We may also define a distance measure for channels acting on a given state $\rho$ called the entanglement fidelity:
\begin{equation}
	F_{\rho}(\mathcal{P},\mathcal{Q}) = f((\mathcal{P}\otimes I)(\ket{\psi}\bra{\psi}), (\mathcal{Q}\otimes I)(\ket{\psi}\bra{\psi})),
\end{equation}
where $\ket{\psi}$ is a purification of $\rho$. This can be used to define the Bures distance between channels:
\begin{equation}
	d_{\rho}(\mathcal{P},\mathcal{Q}) = \sqrt{1-F_{\rho}(\mathcal{P},\mathcal{Q})}.
\end{equation}

In \cite{Beny:2010a}, they introduce the notion of ``$\epsilon$-correctability,'' in which an error is $\epsilon$-correctable if the maximal Bures distance of the noise-recovery composite channel and the identity is bounded:
\begin{equation}
	d(\mathcal{R}\circ\mathcal{N},I) = \max_\rho d_{\rho}(\mathcal{P},\mathcal{Q}) = \sqrt{1-F(\mathcal{R}\circ\mathcal{N},I)} \leq  \epsilon\ .
\end{equation}
In this expression, $F(\mathcal{P},\mathcal{Q})$ is the worst-case entanglement fidelity between two channels $\mathcal{P},\mathcal{Q}$ defined by
\begin{equation}\label{worstcase}
	F(\mathcal{P},\mathcal{Q}) = \min_{\ket{\psi}\in \mathcal{C}^{\otimes 2}}f((\mathcal{P}\otimes I)(\ket{\psi}\bra{\psi}), (\mathcal{Q}\otimes I)(\ket{\psi}\bra{\psi})),
\end{equation}
where the minimization is taken over all pure states in a doubled code subspace. This quantity lower bounds the state fidelity of the original state and the output from $\mathcal{P}$ \cite{Schumacher:1996a}:
\begin{equation}
	F(\mathcal{P},I) \leq f(\rho,\mathcal{P}(\rho)).
\end{equation}
A necessary and sufficient condition for $\epsilon$-correctability for a given noise channel $\mathcal{N}$ and code subspace $\mathcal{C}$ was then given:
\begin{equation}\label{approx}
	PE_i^\dagger E_jP = C_{ij}P + PB_{ij}P,
\end{equation}
where $P$ is the projector onto $\mathcal{C}$, $C_{ij}$ are components of a density matrix, and $d(\mathcal{N},\mathcal{N}+\mathcal{B}) \leq \epsilon$, where $\mathcal{N}(\rho) = \sum_{i,j}\Tr(\rho)C_{ij}\ket{i}\bra{j}$ and $(\mathcal{N}+\mathcal{B})(\rho) = \mathcal{N}(\rho) + \sum_{i,j}\Tr(\rho B_{ij})\ket{i}\bra{j}$. 

The condition (\ref{approx}) leads to a corollary \cite{Brandao:2017irx}: let $\{\ket{\psi_1},\ldots,\ket{\psi_{2^k}}\}$ be an orthogonal set of states in $\mathbb{C}^{2^n}$ such that 
\begin{equation}\label{approx2}
	\braket{\psi_i|E|\psi_j} = C_E\delta_{ij} + \epsilon_{ij}
\end{equation}
for all $i,j$ and any $d$-local operator $E$. Then $\mathcal{C}=\text{span}\{\ket{\psi_1},\ldots,\ket{\psi_{2^k}}\}$ is an $[[n,k,d]]$ $\epsilon$-correctable code with
\begin{equation}
	\epsilon \leq 2^{d+2k}\max_{ij} \epsilon_{ij}^{1/2}
\end{equation}
This implies there exists a recovery channel $\mathcal{R}$ such that:
\begin{equation}
	f(\rho,(\mathcal{R}\circ\mathcal{N})(\rho)) \geq F(\rho,(\mathcal{R}\circ\mathcal{N})(\rho)) \geq 1 - \epsilon^2
\end{equation}

The above corollary is specialized to a system constructed from qubits for concreteness, but it clearly generalizes in a straightforward way. The important thing to note about the error term is that it scales strongly with the code distance and size of the code subspace. Intuitively this makes sense: if we try to correct larger errors or encode more information, the accuracy of the code gets worse. On the other hand, as we will see for the ETH code, $\epsilon$ can also scale very strongly with the system size $n$, so that if $n > k,d$ for $n \gg 1$, the error is small.

\section{ETH as AQECC}\label{ethasaqecc}
Comparing (\ref{approx2}) to (\ref{eth}), we arrive at the observation made in \cite{Brandao:2017irx}: regions of finite energy density in the spectrum of theories that satisfy ETH form $\epsilon$-correctable $[[n,k,d]]$ approximate quantum error correcting codes with
\begin{equation}\label{error}
    \epsilon \leq O(2^{d+2k}e^{S/4}),
\end{equation}
where we have explicitly peeled off the scaling with $d$ and $k$. We now explore this statement from the perspective of approximate error correction. We will henceforth assume that the noise channel we work with is the erasure channel given by (\ref{erasure}). This channel is local, and we therefore expect ETH to hold for erasures on regions smaller than half the total system size. 

Although $\epsilon$-correctability guarantees the existence of a recovery channel, we would like to write down an explicit such channel. \cite{Beny:2010a} and \cite{Beny:2011a} present generic methods to determine a recovery channel in terms of an optimization problem, and we can also write down an explicit recovery map that achieves a state fidelity $f(\rho,\rho') = 1 - O(e^{-S/2})$ between the initial state and recovered state. These maps are discussed in Appendix \ref{expreco}. These maps are somewhat impractical, as the first method requires solving an optimization problem, and the explicit recovery map requires one to determine and diagonalize the matrix $C_{ij}$ in (\ref{approx}). Below, we show that the existence of such maps is sufficient to apply the much more practical theory of universal recovery channels.

\subsection{Universal recovery channel}
We now exhibit a universal recovery channel that allows us to approximately recover a state to good accuracy in the ETH code. The channel is universal in the sense that it does not depend on the input state. To do so, we recast the error correction condition into a statement about relative entropies. Let $\rho,\sigma \in \mathcal{S}(\mathcal{H})$ be density matrices on a Hilbert space $\mathcal{H}$. The relative entropy between the two states is defined by:
\begin{equation}
D(\rho||\sigma) = \begin{cases}
\text{tr}(\rho\log \rho - \rho\log\sigma),& \text{supp }\rho\subseteq \text{supp }\sigma\\
\infty,& \text{ else}
\end{cases}
\end{equation}
Two useful properties of this relative entropy are positivity $D(\rho||\sigma) \geq 0$ with equality if and only if $\rho = \sigma$, and monotonicity under positive, trace-preserving maps $D(\rho||\sigma) \geq D(\Phi(\rho)||\Phi(\sigma))$. These allow us to interpret the relative entropy as a way to measure how ``close'' two states are; the monotonicity property tells us that no quantum channel exists that can increase the distinguishability of two states. The application to error-correction in one direction is immediately obvious: if a recovery map $\mathcal{R}$ exists for a noise channel $\mathcal{N}$, then:
\begin{equation}
D(\rho||\sigma) \geq D(\mathcal{N}(\rho)||\mathcal{N}(\sigma)) \geq D((\mathcal{R}\circ \mathcal{N})(\rho)||(\mathcal{R}\circ \mathcal{N})(\sigma)) = D(\rho||\sigma) .
\end{equation}
So a recovery channel exists only if $D(\rho||\sigma) = D(\mathcal{N}(\rho)||\mathcal{N}(\sigma))$. In \cite{Wilde:2015xoa}, it was shown that the converse is also true: for any two states on a finite-dimensional Hilbert space with non-divergent relative entropy, there exists a recovery map $\mathcal{R}$ that satisfies:
\begin{equation}
D(\rho||\sigma) - D(\mathcal{N}(\rho)||\mathcal{N}(\sigma)) \geq -2\log f(\rho,(\mathcal{R}\circ \mathcal{N})(\rho))
\end{equation}
and $(\mathcal{R}\circ \mathcal{N})(\sigma) = \sigma$.

In \cite{Junge:2015lmb}, they further extend the statement about by showing that $\mathcal{R}$ can be explicitly written down as a universal channel called the twirled Petz map:
\begin{equation}\label{universal}
	\mathcal{R}_{\sigma,\mathcal{N}}(\rho) = \int_{\mathbb{R}}dt\,\frac{\pi}{2}\left(\cosh(\pi t) + 1\right)^{-1}\sigma^{-it/2}\mathcal{P}_{\sigma,\mathcal{N}}\left[\mathcal{N}(\sigma)^{it/2}(\rho)\mathcal{N}(\sigma)^{-it/2}\right]\sigma^{it/2},
\end{equation}
where $\mathcal{P}_{\sigma,\mathcal{N}}$ is the Petz map
\begin{equation}
	\mathcal{P}_{\sigma,\mathcal{N}}(\rho) = \sigma^{1/2}\mathcal{N}^*\left[\mathcal{N}(\sigma)(\rho)\mathcal{N}(\sigma)^{-1/2}\right]\sigma^{1/2}.
\end{equation}
The channel is universal in the sense that it is independent of $\rho$. More precisely, they prove the statement:
\begin{equation}
D(\rho||\sigma) - D(\mathcal{N}(\rho)||\mathcal{N}(\sigma)) \geq -2\log f(\rho,(\mathcal{R}_{\sigma,\mathcal{N}}\circ \mathcal{N})(\rho))
\end{equation}
for arbitrary, possibly infinite, dimension

In doing so, they were also able to show that the error correction condition is robust while preserving the universality of the recovery channel. It isn't too hard to see that if
\begin{equation}
D(\rho||\sigma) - D(\mathcal{N}(\rho)||\mathcal{N}(\sigma)) \leq \epsilon,
\end{equation}
then the universal recovery channel recovers $\rho$ with fidelity
\begin{equation}
f(\rho,(\mathcal{R}_{\sigma,\mathcal{N}}\circ \mathcal{N})(\rho)) \geq 1 - \frac{1}{2}\epsilon.
\end{equation}
By choosing $\sigma = \Pi$ to be the normalized projector onto the code subspace, $\mathcal{R}_{\Pi,\mathcal{N}}$ is manifestly universal for a given channel and choice of code subspace. We may then apply this to our situation: because ETH satisfies (\ref{approx}), we know there exists a recovery channel $\mathcal{R}$ with 
\begin{equation}
	F(\mathcal{R}\circ\mathcal{N},I) \geq 1 - \epsilon^2 \geq 1 - s^{2d+4k}\max_{ij} \epsilon_{ij} = 1 - O(e^{-S/2}).
\end{equation}
We could also use any of the recovery channels from Appendix \ref{expreco} -- their existence is all we need. By the monotonicity of the relative entropy, we then have
\begin{equation}
D(\rho||\Pi) - D(\mathcal{N}(\rho)||\mathcal{N}(\Pi)) \leq D(\rho||\Pi) - D((\mathcal{R}\circ \mathcal{N})(\rho),(\mathcal{R}\circ \mathcal{N})(\Pi)),
\end{equation}
Defining $\rho' = (\mathcal{R}\circ\mathcal{N})(\rho)$, the last term is:
\begin{align}
D((\mathcal{R}\circ \mathcal{N})(\rho),(\mathcal{R}\circ \mathcal{N})(\Pi)) &= \text{tr}[\rho'\log\rho'] - \text{tr}[\rho'\log\Pi'] \\
&= \text{tr}[\rho\log(\rho) - \rho\log(\Pi)] + \delta \\
&= D(\rho||\Pi) + \delta,
\end{align}
where $\delta = O(e^{-S/2})$. A subtlety in the statements above is that we have assumed the relative entropy is not divergent, i.e. $\text{supp }\mathcal{N}(\rho)\subseteq \text{supp }\mathcal{N}(\Pi)$ and $\text{supp }\rho'\subseteq \text{supp }\Pi'$.  Because we chose $\Pi$ to the projector onto the entire code subspace, the support of $\mathcal{N}(\Pi)$ will be the entire mapped code subspace $\mathcal{C}_{\mathcal{N}} = \mathcal{N}(\mathcal{C})$, and naturally $\mathcal{N}(\rho)\in \mathcal{C}_{\mathcal{N}}$. Similarly, it is easy to see that we have $\text{supp }\rho'\subseteq \text{supp }\Pi'$. Therefore, the relative entropies above are well-defined, and we get
\begin{equation}
    D(\rho||\Pi) - D(\mathcal{N}(\rho)||\mathcal{N}(\Pi)) \leq \delta.
\end{equation}
We conclude that we can use $\mathcal{R}_{\Pi,\mathcal{N}}$ as a universal recovery channel that recovers the initial state with fidelity $f(\rho,\rho') \geq 1 - \frac{1}{2}\delta = 1- \mathcal{O}(e^{-S/2})$.

A couple comments are in order: first, the fact that we can use $\mathcal{R}_{\Pi,\mathcal{N}}$ as a good recovery channel is relatively practical, as it manifestly only requires us to know how to implement a particular noise channel and construct $\Pi$. We contrast this with the explicit recovery maps constructed in Appendix \ref{expreco}, which requires that one obtain all the elements of the matrix $C_{ij}$ in (\ref{approx}) or solve an optimization problem. Next, the coefficient appearing in $O(e^{-S/2})$ can be quite large due to the exponential scaling of the bound with $d$ and $k$:
\begin{equation}
	\epsilon^2 \leq s^{2d+4k}\max_{ij}\epsilon_{ij} = O(e^{2d+4k - cn/2}),
\end{equation}
where we have written $S = cn$ as an extensive function of the system size. As noted previously, the scaling with $k$ is not much of an issue, as we usually take $n \gg k$. The scaling with the distance is potentially troublesome, because we have been expecting $d = O(n)$ (the size of the subregion to which ETH applies). Depending on the size of the coefficient $c$, the error can become $O(1)$. However, we expect that bound to be fairly tight in the sense that as long as $n$ is large, we get $e^{(a-c/2)n} \ll 1$ even if $c/2$ is reasonably close to $a$.

Finally, we note that if the conjecture in \cite{Garrison:2015lva} regarding the operators to which ETH applies is correct, we have even stronger correction properties. Not only are we able to correct for local errors, we can also correct for Class I non-local errors as well as Class II errors below a critical energy density. The argument above can then be applied for any other noise channel $\mathcal{N}$, and we can then construct the corresponding $\mathcal{R}_{\Pi,\mathcal{N}}$ as a good recovery channel for those errors. Any system that satisfies ETH therefore has extremely powerful error correcting codes in their spectrum.

\subsection{Chaos as quantum error correction}
A useful information-theoretic perspective for error correction is the one taken in \cite{Beny:2010a}, which is a generalization of the decoupling approach to quantum channels. In this approach, the performance of a code is characterized by how much logical information ``leaks'' to the environment in the noise-plus-recovery channel. The more information leaked, the worse the code performs at protecting from errors. This has a technical definition in terms of a ``complementary channel,'' but we will take the following intuitive notion of this criterion: a code can correct errors exactly if there is no local measurement the environment can do to obtain information about the logical state. That is, given access to an erased subregion, the environment or any other adversary performing local measurements cannot distinguish the global state from any other state in the code subspace. In the framework, we understand approximate QEC as corresponding to states in the code subspace having some amount of distinguishability.

More concretely, we can consider a bipartite tensor factorized Hilbert space $\mathcal{H} = \mathcal{H}_A \otimes \mathcal{H}_{\bar{A}}$ with $|\mathcal{H}_A| < |\mathcal{H}_{\bar{A}}|$. Now introduce a reference system $\mathcal{H}_R$ with $|\mathcal{H}_R| = |\mathcal{H}|$ to represent the environment, and suppose the environment is allowed access to $A$. Let $R$ be a purification of $A\bar{A}$. An equivalent condition for exact error correction is
\begin{equation}\label{mutual}
	I(R:A) = S_R + S_{A} - S_{AR} = 0,
\end{equation}
where $I(A:B)$ is the quantum mutual information that quantifies correlations between systems $A$ and $B$. (\ref{mutual}) is the statement that the environment has no correlations with the subregion $A$, i.e. the errors occurring on $A$ do not leak any information to $R$, which would show up as correlations between the two systems. This is equivalent to requiring:
\begin{equation}
	\rho_{RA} = \rho_R \otimes \rho_A,
\end{equation}
so the systems are entirely decoupled. 

The intuition behind a chaotic system is that the degrees of freedom in the theory are spread out randomly among the entire system; the quantum information of the system is scrambled\footnote{We are intentionally vague as to the technical definition of scrambling here, so that the arguments in this section are general. Rather, we just consider scrambling to refer to some process by which quantum information is delocalized and correlations are suppressed}. Concretely, for a system $\mathcal{H} = \mathcal{H}_A \otimes \mathcal{H}_{\bar{A}}$, we might reasonably model this as by having the reduced state on $A$ carry no quantum information. This is achieved by requiring the state on $A$ to be maximally mixed:
\begin{equation}\label{maxmixed}
\rho_A = \Tr_{\bar{A}}\rho = \frac{1}{\dim \mathcal{H}_A}I_A.
\end{equation}
If the global state is pure $\rho = \ket{\Psi}\bra{\Psi}$ so that $A$ is maximally entangled with $\bar{A}$, then (\ref{maxmixed}) implies (\ref{mutual}), because the monogamy of entanglement ensures that $A$ cannot share any correlations with $R$, automatically forcing $I(R:A) = 0$. We have therefore arrived at the observation that perfectly chaotic (random) systems also provide powerful quantum error correcting codes! In fact, perfectly random scrambling provides the optimal protection against erasure errors \cite{Hayden:2008a,Horodecki:2008,Hayden:2008b}.

This can be understood quite intuitively: randomly scrambling quantum information corresponds to maximally delocalizing information into the entire system, which is precisely the kind of behavior we want in quantum error correction. With so much redundancy, erasing any small region of the system does very little damage to the encoded information; equivalently, any measurements on a small region of the system yield no information. Scrambling essentially functions as an encoder for a random subspace code that is well-protected against erasure errors.

In the context of ETH, we do not have an exact quantum error correction code, but we also did not require that the reduced state on a subsystem be maximally mixed. The systems we consider are not perfectly random, because they cannot actually explore all regions of phase space due to constraints from conservation laws. The conservation of energy carves out a subsystem of phase space that the system is restricted to; this restriction causes reduced states to be approximately thermally entangled rather than maximally mixed, which is a weaker condition. We therefore intuitively understand the error term in the ETH code to be a result of weak, but non-zero, correlations that can exist between a subsystem and the environment. In the thermodynamic limit, the error term vanishes because the equivalence of the canonical and microcanonical ensembles corresponds to the equivalence between thermal and maximally mixed states. 

To see how this functions in practice, suppose we choose an arbitrary state in the code subspace, so that we can write the state as:
\begin{equation}
    \rho = \sum_{ij} c_{ij}\ket{\psi_i}\bra{\psi_j},
\end{equation}
for energy eigenstates $\ket{\psi_i}$ that span the chosen energy band. The expectation value of a local operator $\mathcal{O}$ supported only on $A$ in this state is:
\begin{equation}\label{mixed}
    \braket{\mathcal{O}} = \Tr(\rho\mathcal{O}) = \sum_{ij}c_{ij}\braket{\psi_j|\mathcal{O}|\psi_i} = \braket{\psi_1|\mathcal{O}|\psi_1} + O(e^{-S/2}).
\end{equation}
The choice to use the expectation value in $\ket{\psi_1}$ was arbitrary, and we could have chosen any other eigenstate in the band. The point here is that the variation of the expectation value for an operator obeying ETH in an energy band is exponentially-suppressed. So at the level of observables, an experimentalist cannot determine the global state by performing local measurements on $A$ without exponential measurement accuracy. Correspondingly, the ETH code is can correct for errors up to exponential accuracy. Note that we did not have to restrict to pure states -- we may correct for any state, pure or mixed, in the code subspace. Moreover, we do not even need to know what the state is to correct for errors, by virtue of the universal recovery channel in the previous subsection.

A brief discussion of codes whose reduced subsystems look thermal was given in \cite{Faist:2019ahr} under the name ``thermodynamic codes.'' The approximate nature of these codes is understood to arise due to the fact that energy eigenstates with different energies have different values of the local temperature $T \propto U/n$, where $U$ is the energy of the global energy eigenstate and $n$ is the number of particles/lattice sites. Indeed, it was always possible, in principle, for the environment to learn what the global energy eigenstate is for any finite $n$ and distinguish it from other eigenstates in the code subspace, because the energy eigenvalues in a given energy band are not identical. In the limit $n \to \infty$, the temperature difference for different energy eigenstates vanishes, allowing for arbitrarily accurate error correction. A subtle point to be aware of is that the error in a thermodynamic codes scales as
\begin{equation}\label{errorscale}
\epsilon = O\left(\frac{1}{n}\right),
\end{equation}
whereas for ETH, the error scales as $e^{-O(n)}$. We again stress that ETH really encodes two distinct statements made clear by the subsystem formulation: reduced states of energy eigenstates are exponentially close to a universal state, and the universal state is polynomially close to a thermal state. The thermodynamic code specifically refers to error correction with thermal states, for which ETH also says provide error correction with polynomial accuracy. More generally, the error scaling in (\ref{errorscale}) is a specific example of a more general result on the accuracy of codes that obey $U(1)$-symmetry constraints \cite{Faist:2019ahr, Woods:2019fpy}.

Based on the analysis above, we are led to believe that \emph{all} chaotic systems will contain error correcting codes as a result of their intrinsic scrambling dynamics. If one takes ETH as the working definition of chaos, this is naturally self-evident, but the statement is more general, because it is agnostic to the mechanism of chaos. Any symmetries that the code respects, e.g. energy conservation, will show up as an error term that makes the error correction approximate, because the environment is able to learn about the global state based on differing values of the corresponding conserved charges \cite{Faist:2019ahr,Woods:2019fpy}. ETH is then just a ``kinematic'' statement of the error correcting properties for a nearly-ergodic system that conserves energy; indeed, from the perspective of error correction, we might have guessed that a property like ETH must exist for chaotic systems purely based on the information-theoretic consequences of chaos. 

We close this section by stressing that the error correction properties of ETH do not rely on the above arguments being true, nor does it rely on reduced states of global energy eigenstates looking thermal. Rather, it is a purely a consequence of the ETH ansatz for operator elements, which we take to be the definition of ETH.

\section{Black Holes and Total Bulk Reconstruction}\label{bhandtbr}
We now specialize to the case of holographic CFTs in the AdS/CFT correspondence. As we will see, the error correction properties of these CFTs imply a very strong reconstruction property on the gravity size of the correspondence.

The main concept we use from AdS/CFT is subregion-subregion duality. Given a $d$-dimensional holographic CFT on $S^{d-1}$, this theory is dual to a theory of quantum gravity on AdS$_{d+1}$; we say that the CFT lives on the boundary of the bulk theory of gravity. Subregion-subregion duality makes this statement more precise by identifying which regions of the spacetime are dual to which regions of the CFT. Duality here refers to the existence of an isometry $V$ that maps the bulk to boundary such that
\begin{equation}
    V\mathcal{O}_{\text{bulk}} = \mathcal{O}_{\text{CFT}} V 
\end{equation}
for a bulk operator $\mathcal{O}_{\text{bulk}}$ and boundary operator $\mathcal{O}_{\text{CFT}}$.

\begin{figure}[t!]
	\centering
	\includegraphics[scale=0.4]{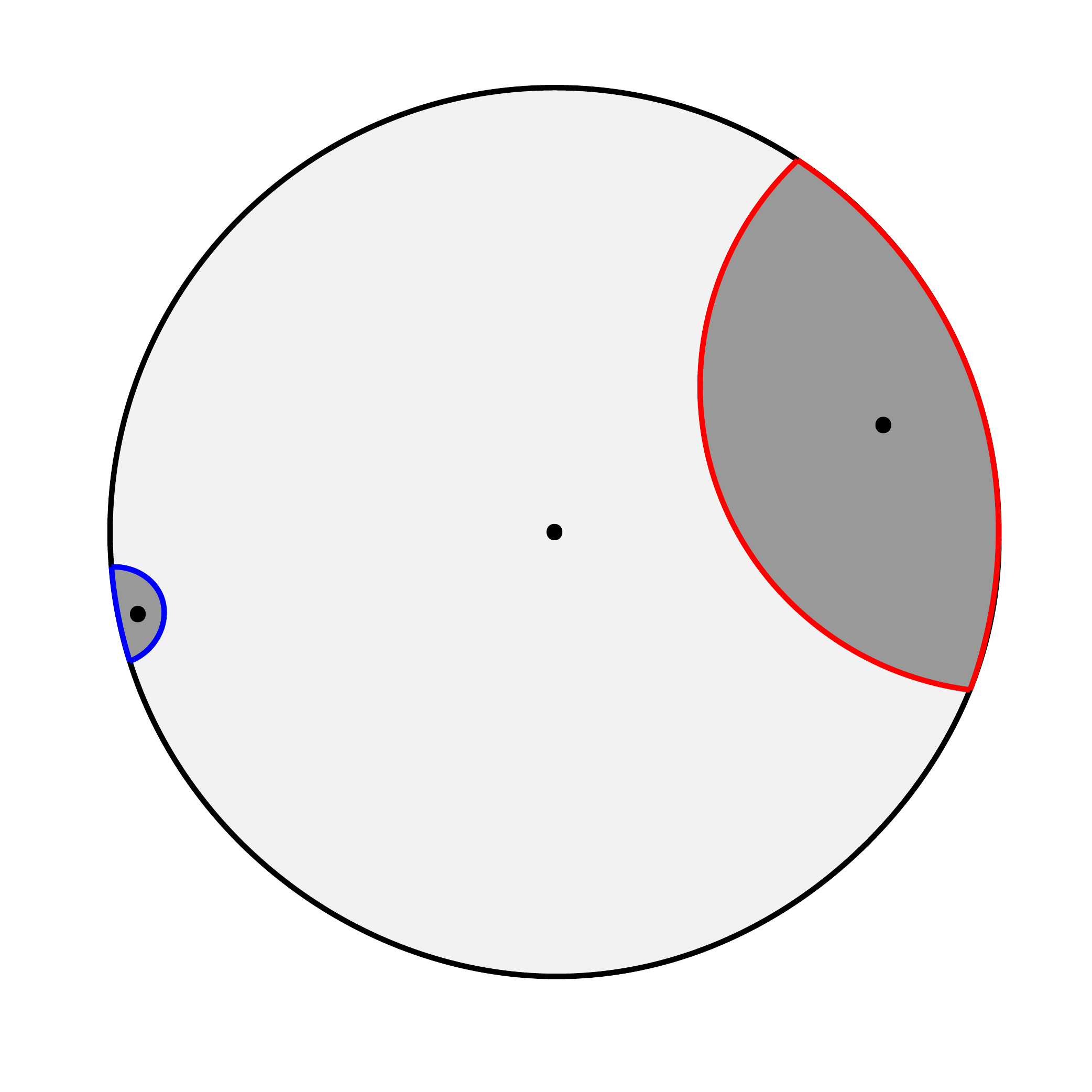}
	\caption{Given a subregion on the boundary, it can be uniquely associated to a dual spacetime region called the ``entanglement wedge'' of the subregion, as shown by the shaded region. An operator in the entanglement wedge, represented by a black dot, has a representation as an operator on the boundary. A larger boundary subregion (red) has an entanglement wedge that goes deeper into the bulk than a smaller one (blue). The operator at the center is well-protected against erasure errors on the boundary; it is unaffected by the deletion of the red or blue region.}
	\label{ewedge}
\end{figure}

Given a region of the CFT with reduced density matrix $\rho_A$, the corresponding dual bulk region is the entanglement wedge $\mathcal{W}[A]$ of $\rho_A$. This is defined by first computing the extremal surface anchored to $\partial A$ and is homologous to $A$. This is the HRT surface, whose area computes the entanglement entropy of $\rho_A$. The entanglement wedge is then defined as the domain of dependence of the surface bounded by the boundary region and the extremal surface, as in Figure \ref{ewedge}.

In \cite{Almheiri:2014lwa} and \cite{Harlow:2016vwg}, it was shown that the mechanism behind the duality can be explained if we treat AdS/CFT as a quantum error correcting code. In this picture, an effective bulk theory acts as the logical data in a code subspace that is encoded in the larger physical space of the CFT. More precisely, any logical operator in the bulk will have a physical representation on a boundary region as a CFT operator if the bulk operator is in the entanglement wedge of the region. These holographic codes are examples of operator algebra quantum error correcting codes -- instead of recovering a full state, we are interested in recovering an algebra of operators. When an erasure occurs on the boundary, the bulk information in the entanglement wedge of the erased region is lost. However, operators deep in the bulk are more strongly protected from erasures, as a subregion must be larger to reach deeper into the bulk, which can also be seen in Figure \ref{ewedge}.

\subsection{ETH and holography}
The relation between ETH and holography is simple: it is expected that holographic CFTs are chaotic, and therefore satisfy ETH. Although this statement has not yet been conclusively proven, the expectation is motivated by the fact that properties of chaotic theories turn out to be also be characteristic of holographic theories. These include large numbers of degrees of freedom (corresponding to ``large $N$''), large central charge, and strong interactions. Explicit non-trivial holographic calculations also provide evidence about the chaotic nature of holographic CFTs.  In \cite{Lashkari:2017hwq}, they compute the entropy of a black hole dual to a holographic CFT in arbitrary dimensions and show that its entropy is in good agreement with the entanglement entropy of the universal ETH density matrix in the low temperature/high energy limit. In \cite{Hikida:2018khg} and \cite{Brehm:2018ipf}, various correlation functions for a 2D CFT on a torus are computed using modular properties, and both the diagonal and off-diagonal elements are shown to agree with the behavior predicted by ETH. Recently, it was shown in \cite{Datta:2019jeo} that typical high energy eigenstates of a 2D CFT even at finite central charge have correlation functions that look thermal, agreeing with the predictions of subsystem ETH.

To make contact with error correction in a CFT context, we follow \cite{Pastawski:2016qrs} in how to interpret code properties in a generic physical system. The length of a code $n$ is corresponds to the size $\dim \mathcal{H}$ of the full Hilbert space $\mathcal{H}$. We imagine that the CFT is placed on a lattice with a finite number of sites $n$, each of which also contains a finite-dimensional Hilbert space $\mathcal{H}_s$. Then the length of the code is the total number of sites on the lattice, $k = \frac{\log \dim \mathcal{C}}{\log\dim\mathcal{H}_s}$, and $d$ is the smallest number of sites that cannot be corrected. We then imagine taking the continuum limit of the lattice system $n \to \infty$. This allows to think about a physical picture where $n$ is the actual system size and $d$ is the size of the smallest region that cannot be corrected.

Now let us interpret the ETH code holographically. We work with the assumption that the encoded logical data is bulk geometry -- this assumption is the essence of the quantum error correction perspective of AdS/CFT and allows us to translate the code properties of the CFT into statements about the bulk. For ETH to apply, we must choose an energy band of width $\Delta E$ that is small relative to the spectrum, but significantly larger than the energy spacing. The code subspace is the set of energy eigenstates $\{\psi_1,\ldots,\psi_{s^k}\}$ that span the band. Although the ETH property is a statement about a single energy eigenstate $\ket{\psi_i}$ and its reduced state $\rho_A = \Tr_{\bar{A}}\ket{\psi_i}\bra{\psi_i}$, the error correction properties apply to any state in the code subspace, as one can see in (\ref{mixed}).

Holographically, an energy eigenstate in the CFT is dual to a black hole in the bulk with ADM mass equal to the energy eigenvalue. Specializing to a pure superposition of energy eigenstates:
\begin{equation}
	\ket{M} = \sum_ic_i\ket{\psi_i}
\end{equation}
This state is dual to a superposition of black hole geometries, where each mass value is given by the corresponding energy eigenvalue. However, because the energies, and hence mass, values are exponentially close, each black hole looks like any other other black hole in the superposition. That is, to exponential accuracy, we can say:
\begin{equation}
	M = \sum_i|c_i|^2E_i \sim E_i.
\end{equation}
The black hole has a definite macroscopic geometry, composed of a set of indistinguishable microstates with energies exponentially close to the macroscopically observed mass. We compare this to the more common case of black hole in a thermal state with definite temperature $\beta^{-1}$:
\begin{equation}
	\rho = \sum_ip_i\rho_i,
\end{equation}
where $p_i = \frac{e^{\beta E_i}}{Z}$ and $\rho_i = \ket{\psi_i}\bra{\psi_i}$. For large system size, the distribution is strongly peaked around the mean energy $M = \sum_ip_iE_i$, which corresponds to the mass of the black hole. The small band spanned by the peak is then the code subspace. The number of microstates in the band is determined by the entropy of the black hole $S = \frac{A}{4G} \sim M^{-1}$. This is how we understand the intuitive notion of ETH as a larger subsystem thermalizing a smaller one.

In the holographic context, the ETH error correction condition says that erasures on less than half of the boundary of a black hole geometry can be approximately corrected, with exponential accuracy that improves with smaller code subspace and smaller erasures. By correct, we mean that if we start with a state $\rho$ and it suffers an erasure error $\mathcal{N}(\rho)$, the full state $\rho$ can be recovered. We do not need to know what the original state was, as we can just apply the universal recovery channel (\ref{universal}) and recover the state. On the bulk side, the effect of the erasure is to erase all the data in the entanglement wedge of the erased region. The ETH code then allows us to, with relative ease, recover all the data in the erased region. In particular, we are able to recover information across the boundary of the cut -- we are not just restoring the information in $\rho_A$, we are restoring the entire $\rho$, including correlations between $\bar{A}$ and $A$ and their corresponding bulk pieces. In other words, we can full state/bulk reconstruction.

Intuitively, this process works because we already mostly know how to patch the state back together. The conditions (\ref{subeth}) and (\ref{ethstate}) tell us that the erased subregion looks like the reduced state on some universal state, which correspondingly looks like a thermal state. Then we can replace the erased subregion with the corresponding universal/thermal subregion. In the process, we stitch the subregion across the dividing surface, which ends up recovering the state to good accuracy.

\subsection{ETH code is not holographic}
\begin{figure}[t!]
	\centering
	\includegraphics[width=\linewidth]{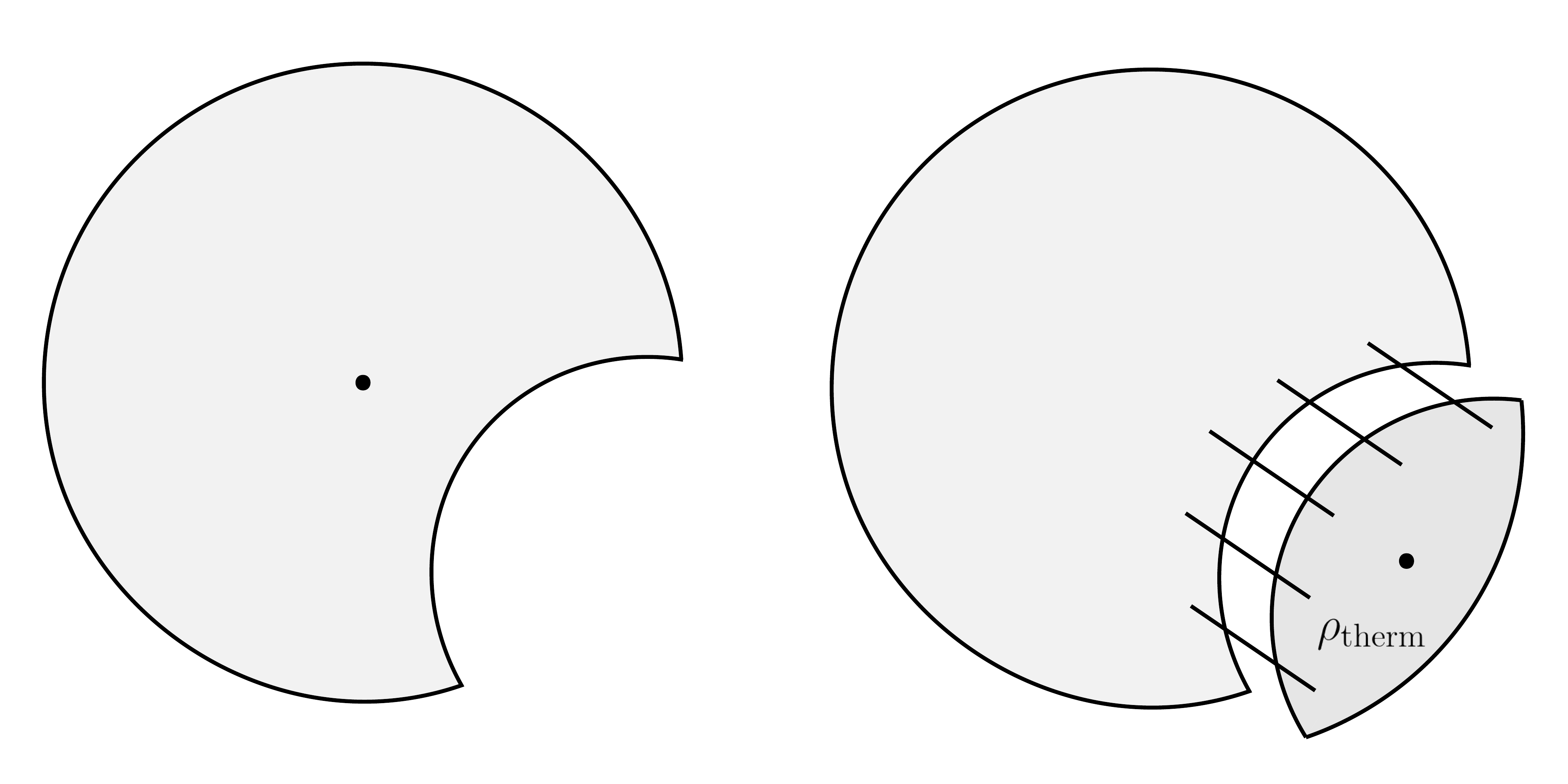}
	\caption{\emph{Left}: In the AdS/CFT code, degrees of freedom deep in the bulk are well-protected from erasure errors, because the size of the erased boundary subregion must be larger in order for its entanglement wedge to reach deeply. The degrees of freedom outside the erased entanglement wedge can be ``reconstructed'' on the remaining boundary region. \newline
	\emph{Right}: In the ETH code, the degrees of freedom in the erased entanglement wedge are restored. It does this by, in a certain sense, patching the erased spacetime with the corresponding spacetime of a thermal state. Moreover, the code does not just restore the data in the erased entanglement wedge, it also restores the correlations with the rest of the spacetime, ``sewing'' the restored entanglement wedge back into place.}
	\label{ETHcorrect}
\end{figure}

We know that the equivalence between the bulk and the boundary in AdS/CFT is only approximate, because we have specialized to an effective bulk theory. Therefore, the error correction must be approximate, rather than exact. Exploration in this direction has yielded some success \cite{Cotler:2017erl}, and it is this direction that formed the original motivation for this work, and it was suggested in \cite{Brandao:2017irx} that the presence of AQECCs in ETH systems is support for the QEC picture of AdS/CFT. However, as we will now show, the two codes are very different, but their combination yields the possibility of total bulk reconstruction against erasure errors for black holes. In particular, we do not expect the error correcting properties of AdS/CFT to be equivalent to the ETH code. The intuition behind their differences is summarized in Figure \ref{ETHcorrect}.

At a fundamental level, the choice of code subspace differs between the two codes. In the ETH code, the code subspace is a small energy band in the spectrum. Holographically, this means that we restrict ourselves to considering a particular black hole spacetime with a well-defined macroscopic geometry. In the AdS/CFT story, we do not have such a restriction, and we expect that there is some form of error correction that holds for generic CFT states dual to an effective bulk theory, even if the boundary state is not constructed from energy eigenstates in a small band. While it is true that the code subspace in the AdS/CFT story is often taken to be small excitations about a given state, the given state need not have the form required for ETH to apply. In particular, we can allow the boundary state to be superposition of energy eigenstates that are not exponentially close, corresponding to a superposition of macroscopically distinct geometries that form a multi-center black hole.\footnote{In fact, there has been some work on a conjecture that multi-center black holes are described by glassy physics. These systems are special for precisely the reason that they do not obey ETH, rather they exhibit many-body localization.} 

The code subspace defines a code -- because they differ, we conclude that their correction properties should differ, and indeed, the actual correction mechanism is very different. The ETH code is the ``strongest" type of error correction, because it recovers a full state. This is not the case in the AdS/CFT story, which is defined by the much more nuanced theory of operator algebra correcting codes, where one instead asks only to recover a subalgebra of operators. This more complicated structure is necessary to capture the complexities of the AdS/CFT duality, e.g. operators deeper in the bulk are more strongly protected than those near the boundary. It is also worth noting that the ETH code only has non-perturbative corrections in the system size, while the AdS/CFT code is expected to have both perturbative and non-perturbative corrections in $N$.

At the level of the code mechanism, the ETH code is simpler because it is fundamentally not holographic. To reiterate, the ETH code is a statement of the error correction properties in a chaotic system, and holographic theories only form a subset of such systems. The claim that an erased subregion on the boundary can be restored is a statement that is purely within the confines of the CFT. We can only extend this to a statement about reconstructing the bulk if the CFT is also holographic. In \cite{Harlow:2016vwg}, this distinction is particularly sharp, because the ETH code corresponds to an approximate version of ``conventional quantum erasure correction,'' which is shown to lack the structure needed to describe a holographic encoding. In particular, the structure of the entropy of the boundary state is insufficient, and the entanglement structure of the theory that allows full state recovery is too restrictive.

\subsection{Complementary bulk reconstruction}
\begin{figure}[t!]
	\centering
	\includegraphics[width=\linewidth]{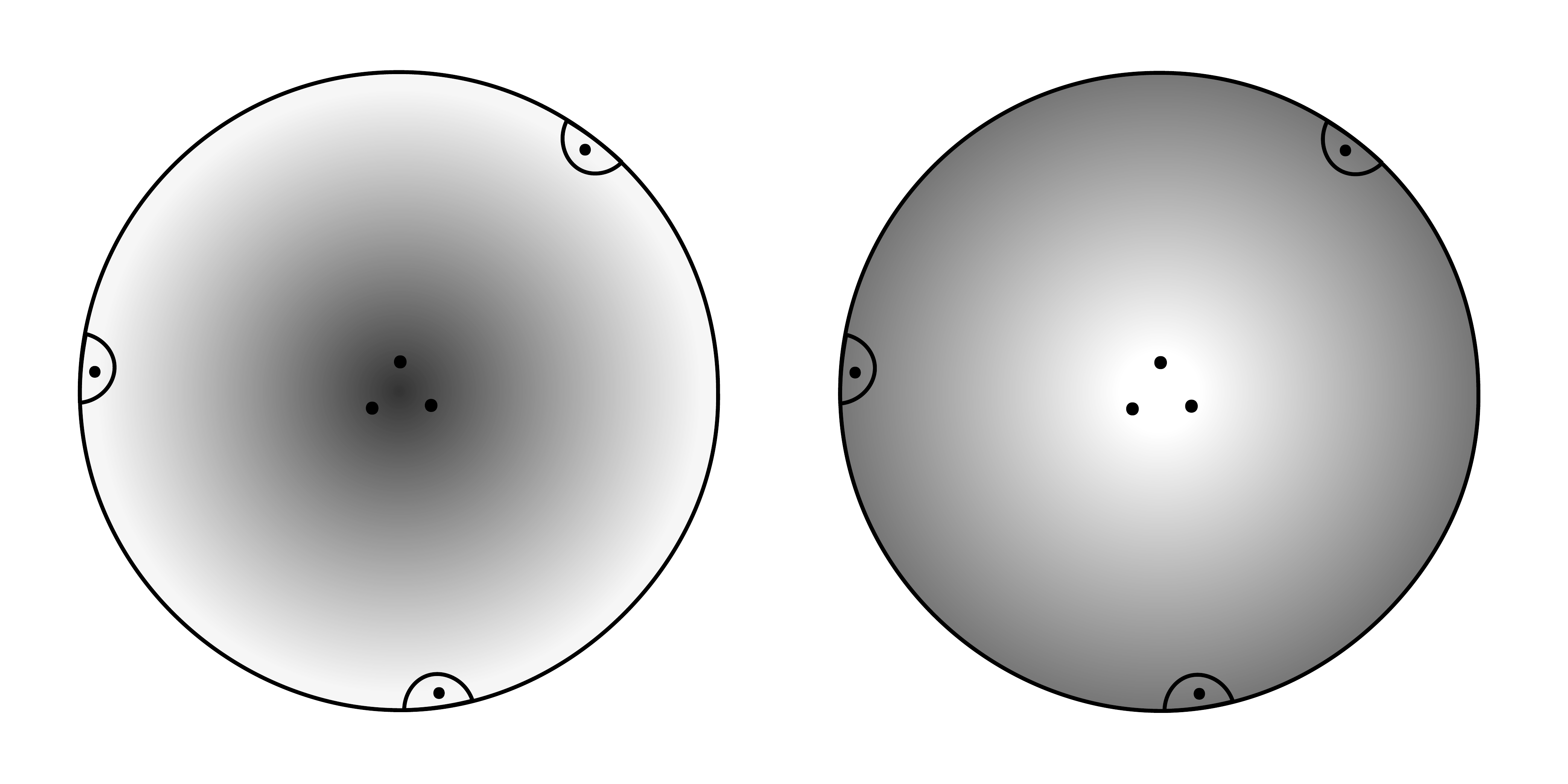}
	\caption{The ETH code, on the left, performs very well for small boundary regions, corresponding to the UV degrees of freedom in the bulk that are near the boundary. As we move inwards into the bulk, the size of the error, represented by the darkening gradient, grows. In the AdS/CFT code on the right, these degrees of freedom are poorly protected because of how shallowly they lie in the bulk. On the other hand, degrees of freedom in the center may not be recovered to good accuracy in the ETH code, but are strongly protected in the AdS/CFT code. In this way, the two codes complement each other by covering for their regions of weak recovery, providing an extremely robust protection for the bulk geometry.}
	\label{compreco}
\end{figure}

While the ETH code may not be intrinsically holographic, its simplicity affords it extra power. The AdS/CFT code tells us how to reconstruct operators that have support on the entanglement wedge of the unerased region of the boundary. Operators deep in the bulk are well-protected because they can be represented on a large class of entanglement wedges, and correspondingly, operators near the boundary are very weakly-protected. On the other hand, the ETH code is explicit reconstruction of the data in the erased region. Moreover, because the entire state is recovered, the code is restoring the entire spacetime of the erased entanglement wedge. This is very powerful, because in defining the separation between the entanglement wedge of $A$ and $\bar{A}$, we had to introduce the geometric Ryu-Takayanagi surface. There are degrees of freedom that live on the RT surface which make generic reconstruction difficult, but the ETH code is nonetheless able to connect the two spacetimes with the (approximately) correct correlations across the RT surface. The ETH code therefore allows for total bulk reconstruction, as shown in Figure  \ref{compreco}.

The AdS/CFT code complements the ETH code in the region where the ETH code begins to have poorer performance, and the converse is true as well. The accuracy of the ETH code decreases with distance, so accuracy is improved for errors on small boundary regions. The entanglement wedge of a small boundary region is shallow in the bulk, so the operators in the wedge are not well-protected in the AdS/CFT code. In contrast, operators deep in the bulk are very well-protected in the AdS/CFT code, but the boundary state dual to an entanglement wedge that goes deep into the bulk is very large, so the accuracy of the ETH code is decreased\footnote{Because the error is still exponential in the system size (\ref{approx2}), even a relatively small, but finite, positive difference in the arguments of the exponentials gives good accuracy. In other words, the code is still very effective for reasonably large distances and code subspace dimensions.}. We see that for states and errors where both codes apply, the bulk geometry is extremely robust against errors due to the complementary regions of good performance of the AdS/CFT and ETH codes. However, due to our lack of precise understanding of the error term in the ETH code, it seems possible that there can be a kind of ``no-mans-land'' where the error in the ETH code is $O(1)$, even though the erased subregion is less than half the system size.

We emphasize here that existence of a recovery channel has deep implications for the non-local nature of the quantum information in the theory. The information contained in the entanglement wedge, while dual to a particular subregion on the boundary, is spread out among the entire system. Note that this statement is not in contention with the existing story of subregion-subregion duality -- it is just a property of random encoding, independent of holography. To see how this works, it is convenient to instead pass to the Heisenberg picture, where operators evolve according to a dual noise channel:
\begin{equation}
	\mathcal{N}^\dagger(X) = \sum_i E_i^\dagger XE_i.
\end{equation}
It is easy to see from this definition that
\begin{equation}
	\Tr(\mathcal{N}(\rho)X) = \Tr(\rho\mathcal{N}^\dagger(X)).
\end{equation}
Correspondingly, the error correction condition can be phrased as
\begin{equation}
	P(\mathcal{R}\circ\mathcal{N})^\dagger(X) P = PXP + \epsilon.
\end{equation}
It is simple to argue that the dual of the channel that erases $A$ maps all operators supported on $A$ to ones that only have support on $\bar{A}$ \cite{Pastawski:2016qrs}:
\begin{equation}
	\mathcal{N}^\dagger(X) = I_A\otimes X'_{\bar{A}},
\end{equation}
from which it is easy to see that recovered operator $(\mathcal{R}\circ\mathcal{N})^\dagger(X) = (\mathcal{N}^\dagger \circ \mathcal{R}^\dagger)(X)$ only has support on $\bar{A}$. This perspective makes very clear the point that the logical information on $A$ is really dispersed around the entire system.

\section{Discussion}\label{disc}
In this paper, we have examined ETH from an error correction standpoint, and applied to analysis to study holographic CFTs. In the process, we found that black holes with well-defined mass are robust against erasure errors to the extent that it is possible to approximately fully reconstruct the bulk against such errors, because the ETH code is distinct and more powerful than the AdS/CFT code. More generally, we argued that this is nothing more than an example of a more general idea that all chaotic theories contain error correcting codes in their spectrum due to their intrinsic scrambling dynamics. That we can interpret ETH as a quantum error correcting code has greater generality than what we explored in the context of holography. We now discuss some connections to other work possible avenues for the future.

\subsection{State distinguishability in black holes}
One of the defining characteristics of ETH is that the energy eigenstates in the chosen band of energy are indistinguishable from each other when an observer can only perform measurements on a subregion less than half the size of the full system. While we made very general arguments to this effect by appealing to the exponentially-suppressed energy spacing and intuitive notions of chaos and ETH, it is interesting to examine the question more carefully for specific systems of interest.

A more quantitative analysis of state distinguishability for black holes in AdS/CFT was done in \cite{Bao:2017guc}, phrased in terms of the Holevo information. A precise statement of the problem is as follows: recall that a black hole of inverse temperature $\beta^{-1}$ in AdS/CFT is dual to a thermal ensemble
\begin{equation}
	\rho = \sum_ip_i\rho_i,
\end{equation}
where $\rho_i = \ket{\psi_i}\bra{\psi_i}$ are microstates (energy eigenstates), and $p_i = \frac{e^{-\beta E_i}}{Z}$. Choosing a small band of energies centered on the mass of the black hole, ETH tells us that we can compute expectation values using a microstate $\rho_i$ in the band to good accuracy, despite the fact that $\rho \neq \rho_i$. State distinguishability is the question of how much we can learn about whether the global state is $\rho$ or a particular microstate $\rho_i$. If we restrict measurements to a subregion $A$, the Holevo information is the ensemble-averaged relative entropy of $\rho$ with $\rho_i$:
\begin{equation}
	\chi(A) \equiv \sum_ip_iD(\rho_A||\rho_{i,A}) = S(\rho_A) - \sum_i p_i S(\rho_{i,A}).
\end{equation}
This quantity is useful because it upper bounds the information we can gain from measurements on $A$ to learn about the global state. It monotonically increases with $A$, starting from 0 when states are not distinguishable at all (the states are maximally mixed), to $S_{bh} = S(\rho)$ when $A$ covers the whole boundary (all the $\rho_{i,A} = \rho_i$ are pure, and hence orthogonal). More precisely, it upper bounds the mutual information between the distribution of measurement outcomes and the distribution of the initial ensemble -- this is called the ``accessible information.''

Following their normalization scheme, the radius of a ball-shaped region on the boundary is $\ell_A = \pi/2$ when $A$ covers half the boundary and $\ell_A = \pi$ when it covers the full boundary. By using the AdS/CFT duality, they showed that in $d = 2$ CFTs, microstates are totally indistinguishable for $\ell_A < \pi/2$, monotonically increasingly distinguishable for $\pi/2 \leq \ell_A < \ell_{\text{crit}}$ for a critical size $\ell_{\text{crit}}$, and completely distinguishable for $\ell_{\text{crit}} \leq \ell_A$. For $d > 2$, it was found that microstates are indistinguishable for $\ell_A < \pi - \ell_{\text{crit}}$ and distinguishable for $\ell_A > \ell_{\text{crit}}$, with some unknown, but increasing behavior in $\pi - \ell_{\text{crit}} \leq \ell_A \leq \ell_{\text{crit}}$. 

Their results agree with our analysis: ETH says that measurements for $\ell_A < \pi/2$, viewing them as local errors, will return results consistent with a thermal average up to exponential corrections, but the energy spacing itself is exponentially suppressed, so the microstates are indistinguishable. We can also interpret their results from the perspective of quantum error correction. Instead of an experimenter performing measurements on the region $A$, we have the environment doing the measurements so that information ``leaks'' out to it. The code is good if the environment cannot get much information about the global state, as in (\ref{mutual}). Therefore, we expect that so long as the ETH code holds to good accuracy, microstates will be indistinguishable. We can think of (\ref{mutual}) as playing the role of the accessible information; when the mutual information is 0, the code is working well, and as it departs from 0, the code begins to fail.

The ETH code is approximate outside of the thermodynamic limit, so the microstates will have some small distinguishability features; the reason this does not show up in the analysis in \cite{Bao:2017guc} is because their analysis is only to leading order -- including corrections will naturally perturb their results to make microstates minutely distinguishable. However, their results are can be useful as a leading order analysis of how the ETH code begins to fail as the $\ell_A \to \pi/2$. In particular, we can see that the ETH code fails quite badly and very quickly once we push the limits of its applicability. A more detailed analysis that keeps track of higher order terms would be reveal how the ETH code begins to fail as $d \to n/2$. For example, such an analysis would allow us to better understand the potential no-mans-land in the ETH code where the error is $O(1)$, which may be compared to the no-mans-lands in the AdS/CFT code \cite{Harlow:2016vwg}. It would also be interesting to see how and to what extent the error terms in the ETH code match or diverge from the expected AdS/CFT quantum corrections.

\subsection{Chaos, QEC, and QG}
We stress that the analysis performed in this work is really a statement about quantum chaos, modeled by ETH, and quantum error correcting codes. We have specialized to holographic CFTs so that we could interpret the error correction properties as statements about gravity. However, the code properties should hold true in any system that satisfies ETH, including many-body or condensed matter systems where the interest is primarily in understanding ETH as a thermalization mechanism. Such a perspective has already been applied successfully in the context of random circuits, e.g. \cite{Choi:2019nhg}, and we hope that the analysis in this work will also prove useful for future work in these directions. As we saw, ETH can be interpreted as a ``kinematic'' statement that falls out of the expectation that a chaotic theory should have error correcting properties.

\begin{figure}[t!]
	\centering
	\includegraphics[scale=0.4]{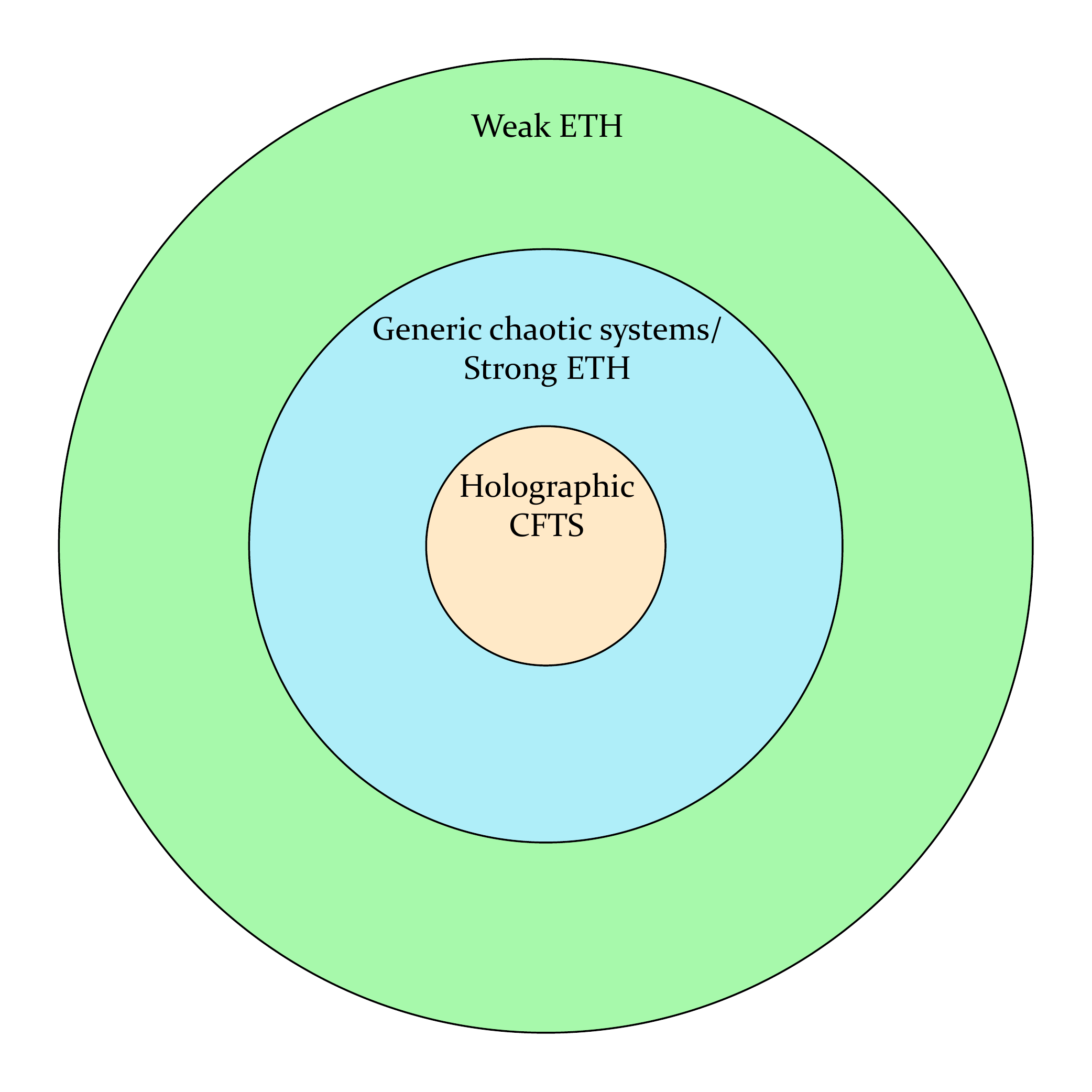}
	\caption{The hierarchy of ETH applicability. Our analysis has concerned the intermediate tier of systems that satisfy strong ETH, in which all eigenstates give thermal expectation values. We suspect that the restriction to just holographic CFTs and the extension to systems that satisfy weak ETH can be understood by analyzing quantum error correcting code properties.}
	\label{hierarchy}
\end{figure}

Quantum error correction may be a useful perspective in gaining a deeper understanding of ETH in general. In our analysis, we assumed that all energy eigenstates for a given band in the spectrum obey ETH; this is usually called \emph{strong} ETH. It is true that fluctuations are exponential in $n$, the Hilbert space size is also exponential in $n$, so it is possible that the spectrum contains an exponentially small number of states that do not obey ETH. This is called \emph{weak} ETH \cite{Biroli:2010a}. The distinction is important because even integrable systems are expected to have behavior that is qualitatively similar to weak ETH, in the sense that the vast majority of states, so-called ``typical'' states, exhibit thermal behavior at the level of operator expectation values \cite{Mueller:2015a}. It may be possible to analyze the boundary separating weak and strong ETH in the context of QEC, as their differences will show up as quantitative differences in their code properties. Along similar lines, we although we focused on holographic CFTs, none of our analysis actually used the holographic property of a CFT. The condition(s) required for a holographic dual may show up as an enhancement of code properties. The general hierarchy can be seen in Figure \ref{hierarchy}.

More generally, we would like to more deeply understand the connection between chaos and quantum error correction. In this work, we have examined chaos from a qualitative perspective, augmented with perspectives from quantum information and evidence from recent work in scrambling. It would be interesting to connect quantitative aspects of chaos, such as the decay of out-of-time-order correlators (OTOC) and operator growth, to quantum error correction \cite{Sonner:2017hxc,Foini:2018sdb,Nie:2018dfe,Zhuang:2019jyq,deBoer:2019kyr, Anous:2019yku, Jensen:2019cmr}. By extension, we can then apply such results to understanding holographic CFTs and their duals, as part of a general program of applying tools and concepts from quantum information and error correction to understanding quantum gravity. Chaos appears to be a necessary, but insufficient condition for a theory to have a holographic dual, so a deeper understanding of the chaos-QEC-CFT relationship will likely provide insight into the general holographic sufficiency conditions. Given how ubiquitous (approximate) error correcting codes seem to be, it seems very likely that there remains a lot of mileage left to get out of the QEC perspective on AdS/CFT. 

Studying quantum gravity can also provide insights into quantum error correction. Motivated by understanding AdS/CFT better, recent work, e.g. \cite{Jahn:2019nmz,Kang:2018xqy,Harris:2018jfl}, has produced interesting results for understanding holographic (operator algebra) codes in general. There may be questions that are more easily answered on the gravity side than the CFT side, and translating those answers into statements about properties of the CFT may then give insight into error correction.

\subsection{The curse of finite dimensions}
The most glaring limitation of this work is that we have worked with finite-dimensional Hilbert spaces, trading reality for simpler calculations and interpretations. While this may be somewhat excusable for a system of qubits, it is clearly insufficient if we wish to analyze CFTs and their holographic duals. It would be interesting and useful to try to generalize our analysis to infinite dimensions. While we expect that our work should generalize, infinite-dimensional error-correcting codes are tricky to work with \cite{Beny:2009a} and their approximate cousins even moreso, although there has been some recent progress \cite{Faist:2019ahr,Woods:2019fpy, Kang:2018xqy}. In particular, we expect that the qualitative story of chaotic theories containing QECCs and, by extension, complementary bulk reconstruction will remain unchanged.

\acknowledgments
We thank Chris Akers, Nick Hunter-Jones, and Beni Yoshida for useful and interesting conversations. NC thanks the YITP at Kyoto University for hospitality during its "Quantum Information and String Theory 2019" workshop, where many of the conversations were held and part of this work was completed. NB is supported by the National Science Foundation under grant number 82248-13067-44-PHPXH, by the Department of Energy under grant number DE-SC0019380, and by New York State Urban Development Corporation - Empire State Development - contract no. AA289.

\appendix
\section{Some recovery channels}\label{expreco}
We first exhibit an explicit recovery map that recovers a state in a code satisfying the ETH ansatz (\ref{eth}) with state fidelity $f = 1 - O(e^{-S/2})$. This recovery map is identical to the recovery map used in textbook proofs \cite{Lidar:2013ed} of the sufficiency of the Knill-Laflamme error correction conditions. Given a set of error elements $\{E_i\}$, we are free to choose a different set of error elements $\{F_i\}$ related to the original by a unitary transformation:
\begin{equation}
F_i = \sum_j u_{ij}E_j.
\end{equation}
It is easy to show that $\{F_i\}$ describe the same noise channel as the original error elements. Choose the unitary matrix $u_{ij}$ to be one that diagonalizes the $C_{ij}$ matrix in (\ref{approx}) so that:
\begin{equation}
\braket{\psi_m|F_i^\dagger F_j|\psi_n} = d_i\delta_{ij}\delta_{mn} + \epsilon_{ij(mn)},
\end{equation}
where $\sum_i d_i = 1 - \sum_i\epsilon_{ii}$ by the assumption that the noise channel is trace-preserving. Then define a recovery channel by the elements:
\begin{equation}\label{recovery}
R_n = \frac{1}{\sqrt{d_n}}PF_n^\dagger = \frac{1}{\sqrt{d_n}}\sum_m \ket{\psi_m}\bra{\psi_m}F_n^\dagger,
\end{equation}
where $P = \sum_m \ket{\psi_m}\bra{\psi_m}$ is the projector onto the code subspace in the energy eigenbasis. It is straightforward to show $R = \sum_nR_n^\dagger R_n$ is a projector onto the subspace of states spanned by all the possible states reached by acting with the error elements on states in the code subspace. Therefore we can take the $R_n$ to define a valid channel, with the possibility that we need to include the projector onto the orthogonal subspace $I - R$, which naturally does not affect the recovery channel. Expanding the original state in the energy eigenbasis $\rho = \sum_{mn}c_{mn}\ket{\psi_m}\bra{\psi_n}$, the state after the noise and recovery channels is
\begin{align}
(\mathcal{R}\circ\mathcal{N})(\rho) &= \sum_{ij}R_i F_j\rho F_j^\dagger R_i^\dagger \\
&= \sum_{mnkl}\sum_{ij}\frac{1}{d_i}c_{mn}\ket{\psi_k}\bra{\psi_k}F^\dagger_iF_j\ket{\psi_m}\bra{\psi_n} F^\dagger_j F_i\ket{\psi_l}\bra{\psi_l},
\end{align}
where the the first summation runs over the dimension of the code subspace, and the second summation runs over the number of operation elements. Now we apply the ETH condition:
\begin{align}
(\mathcal{R}\cdot \mathcal{N})(\rho) &= \sum_{mnkl}\sum_{ij}\frac{1}{d_i}c_{mn}\left[d_i\delta_{ij}\delta_{km} + \epsilon_{ij(km)}\right]\left[d_j\delta_{ji}\delta_{nl} + \epsilon_{ji(nl)}\right]\ket{\psi_k}\bra{\psi_l} \\
&= \sum_{mnkl}\sum_{ij}\frac{1}{d_i}c_{mn}[d_id_j\delta_{ij}\delta_{ji}\delta_{km}\delta_{nl} + d_i\delta_{ij}\delta_{km}\epsilon_{ji(nl)} +  d_j\delta_{ji}\delta_{nl}\epsilon_{ij(km)} \\
& \nonumber\hspace{23.5em}+ \epsilon_{ij(km)}\epsilon_{ji(nl)}]\ket{\psi_k}\bra{\psi_l} \\
&= \sum_{mn}\sum_i d_ic_{mn}\ket{\psi_m}\bra{\psi_n} + \sum_{mnl}\sum_ic_{mn}\epsilon_{ii(nl)}\ket{\psi_m}\bra{\psi_l} \\
& \nonumber\hspace{5em} + \sum_{mnk}\sum_ic_{mn}\epsilon_{ii(nl)}\ket{\psi_k}\bra{\psi_n} +  \sum_{mnkl}\sum_{ij}\frac{1}{d_i}c_{mn}\epsilon_{ij(km)}\epsilon_{ji(nl)}\ket{\psi_k}\bra{\psi_l} \\
&= \rho'.
\end{align}
We recognize the first term as nearly the original state $\left(\sum_{i}d_i\right)\rho = (1-O(e^{-S/2}))\rho$. The other terms are the error terms, where the $\epsilon$ terms are order $O(e^{-S/2})$, and therefore suppressed by the error parameter, as expected. The trace distance of this state to the input is clearly:
\begin{equation}
\frac{1}{2}||\rho - \rho'||_1 = O(e^{-S/2}).
\end{equation}
An application of the Fuchs-van de Graaf inequality then tells us:
\begin{equation}
f(\rho,\rho') \geq 1 - \frac{1}{2}||\rho - \rho'||_1 = 1 - O(e^{-S/2}).
\end{equation}
We must be careful because the terms multiplying the $e^{-S/2}$ term can be quite large on their own, as we see in the error term for (\ref{approx2}). We really do need the length of the code to be much larger than the size of the code subspace in order for the code to be good. Note however that this channel is somewhat impractical, because it requires one to determine the diagonalizing unitary $u_{ij}$, which requires determining the entire matrix $C_{ij}$.

We now review the procedure for constructing a near-optimal recovery channel presented in \cite{Beny:2010a,Beny:2011a}. To so, we first must introduce the theory of complementary channels. Given a generic quantum channel $\mathcal{N}$, we can write it using a Stinespring dilation:
\begin{equation}
    \mathcal{N}(\rho) = \Tr_E\left[V (\rho \otimes \ket{0}\bra{0})V^\dagger\right],
\end{equation}
where we have extended the existing Hilbert space with a another one $\mathcal{H}_E$ representing the environment, and $V$ is an isometric mapping to the extended Hilbert space. The complementary channel is obtained by instead tracing out the input system $A$ and leaving the environment:
\begin{equation}
    \widehat{\mathcal{N}}(\rho) = \Tr_A\left[V (\rho \otimes \ket{0}\bra{0})V^\dagger\right].
\end{equation}
The key application to quantum error correction is the theorem proved in \cite{Beny:2010a}:
\begin{equation}
    \max_\mathcal{R}{F}(\mathcal{R}\circ\mathcal{N},\mathcal{M}) = \max_\mathcal{R'}{F}(\widehat{\mathcal{N}},\mathcal{R}'\circ\widehat{\mathcal{M}}),
\end{equation}
where is $F$ is the worst-case entanglement fidelity. For error correction, we take $\mathcal{M} = I$, so that we get a statement about the maximum worst-case fidelity that can be achieved for a recovery channel $\mathcal{R}$. Moreover, in the case that $\widehat{\mathcal{M}}^2 = \widehat{\mathcal{M}}$, as is clearly the case for error correction, there is a nice bound:
\begin{equation}
    F(\mathcal{R}\circ\mathcal{N},\mathcal{M}) \geq F(\widehat{\mathcal{N}},\widehat{\mathcal{N}}\circ\widehat{\mathcal{M}}).
\end{equation}
The goal is to determine a recovery channel that achieves this behavior. The worst-case entanglement fidelity can be written as a maximin problem:
\begin{equation}
    F(\widehat{\mathcal{N}},\mathcal{R}'\circ\widehat{\mathcal{M}}) = \min_\rho \max_{||A|| \leq 1}\text{Re}\ g_{\rho}(A,U'),
\end{equation}
where $A$ is an operator from the input of $\mathcal{R}$ to the environment-extended Hilbert space, $U'$ is a dilation satisfying $\widehat{N}(\rho) = \Tr_K\left[U'(\rho \otimes \ket{0}\bra{0})U'^\dagger\right]$ for some extension $\mathcal{H}_K$, and $g_{\rho}$ is a bilinear function defined as follows:
\begin{equation}
g_{\rho}(U,U') = \braket{\psi_\rho|V_{\mathcal{M}}^\dagger(I_B'\otimes U'^\dagger_{E'\Tilde{E}})(U_B\otimes I_E)V_{\mathcal{N}}|\psi_\rho},
\end{equation}
where $V_{\mathcal{N}},\ V_{\mathcal{M}}$ are the dilation maps for $\mathcal{N}$ and $\mathcal{M}$, and $\ket{\psi_{\rho}}$ is an arbitrary purification of $\rho$. Notationally, the subscripts indicate the support of the operators: $B$ and $B'$ are the target spaces for $\mathcal{N}$ and $\mathcal{M}$, respectively; $E$ and $E'$ are the extended environmental systems for the target spaces of $V_{\mathcal{N}}$ and $V_{\mathcal{M}}$, respectively; and finally $\Tilde{E}$ is the auxiliary space introduced so that $\mathcal{R}(\rho) = \Tr_{\Tilde{E}}(U\rho U^\dagger)$, and is acted on at the end by sandwiching it between $\ket{0}$'s. No operators act on the purifying subsystem.

The goal is then to find a saddle point $(\rho_0,A_0)$ such that
\begin{equation}
    \text{Re}\ g_{\rho_0}(A_0,U') = \min_\rho \max_{||A|| \leq 1}\text{Re}\ g_{\rho}(A,U')
\end{equation}
and
\begin{equation}
    \text{Re}\ g_{\rho_0}(A_0,U') = \max_{||A|| \leq 1}\min_\rho \text{Re}\ g_{\rho}(A,U').
\end{equation}
After finding the saddle point, one then defines a completely-positive, trace non-increasing map
\begin{equation}
    \mathcal{S}(\rho) = \Tr_K\left[A_0(\rho \otimes \ket{0}\bra{0})A_0^\dagger\right],
\end{equation}
and completes it to a trace-preserving map:
\begin{equation}
    \mathcal{R}_g(\rho) = \mathcal{S}(\rho) + \Tr\left[\rho - \mathcal{S}(\rho)\right]\tau,
\end{equation}
for some arbitrary state $\tau$. Then one can see:
\begin{align}
F(\mathcal{R}_g\circ\mathcal{N},\mathcal{M}) &\geq F(\mathcal{S}\circ \mathcal{N},\mathcal{M}) \\
&= \min_\rho \max_{U'}\text{Re}\ g_{\rho}(A_0,U') \\
&= \max_{U'}\min_\rho \text{Re}\ g_{\rho}(A_0,U')\\
&\geq \min_\rho \text{Re}\ g_{\rho}(A_0,U') \\
&= \text{Re}\ g_{\rho_0}(A_0,U') \\
&= F(\widehat{\mathcal{N}},\widehat{\mathcal{N}}\circ\widehat{\mathcal{M}}).
\end{align}
We may therefore use $\mathcal{R}_g$ as a channel that performs good recovery. Unfortunately, the the channel is relatively impractical, as one must determine the necessary saddle point $(\rho_0,A_0)$. The problem somewhat simplifies when we specialize to the case of $\mathcal{M} = I$ for error correction but a generic optimization problem still remains \cite{Beny:2011a}.


\begin{thebibliography}{99}
\bibitem{Almheiri:2014lwa} 
A.~Almheiri, X.~Dong and D.~Harlow,
``Bulk Locality and Quantum Error Correction in AdS/CFT,''
\href{https://dx.doi.org/10.1007/JHEP04(2015)163}{\emph{JHEP} {\bf 1504} (2015) 163},
\href{https://arxiv.org/abs/1411.7041}{\tt arXiv:1411.7041 [hep-th]}.

\bibitem{Harlow:2016vwg}
D.~Harlow,
``The Ryu–Takayanagi Formula from Quantum Error Correction,''
\href{https://doi/10.1007/s00220-017-2904-z}{\emph{Commun.\ Math.\ Phys}.\  {\bf 354} (2017) no. 3, 865-912},
\href{https://arxiv.org/abs/1607.03901}{\tt arXiv:1607.03901 [hep-th]}.


\bibitem{Pastawski:2015qua}
F.~Pastawski, B.~Yoshida, D.~Harlow and J.~Preskill,
``Holographic quantum error-correcting codes: Toy models for the bulk/boundary correspondence,''
\href{https://dx.doi.org/10.1007/JHEP06(2015)149}{\emph{JHEP} {\bf 1506} (2015) 149}
\href{https://arxiv.org/abs/1503.06237}{\tt arXiv:1503.06237 [hep-th]}.


\bibitem{Akers:2018fow}
C.~Akers and P.~Rath,
``Holographic Renyi Entropy from Quantum Error Correction,''
\href{https://dx.doi.org/10.1007/JHEP05(2019)052}{\emph{JHEP} {\bf 1905} (2019) 052},
\href{https://arxiv.org/abs/1811.05171}{\tt arXiv:1811.05171 [hep-th]}.


\bibitem{Bao:2018pvs}
N.~Bao, G.~Penington, J.~Sorce and A.~C.~Wall,
``Beyond Toy Models: Distilling Tensor Networks in Full AdS/CFT,''
\href{https://arxiv.org/abs/1812.01171}{\tt arXiv:1812.01171 [hep-th]}.


\bibitem{Verlinde:2012cy}
E.~Verlinde and H.~Verlinde,
``Black Hole Entanglement and Quantum Error Correction,''
\href{https://dx.doi.org/10.1007/JHEP10(2013)107}{\emph{JHEP} {\bf 1310} (2013) 107}
\href{https://arxiv.org/abs/1211.6913}{\tt arXiv:1211.6913 [hep-th]}.


\bibitem{Almheiri:2018xdw}
A.~Almheiri,
``Holographic Quantum Error Correction and the Projected Black Hole Interior,''
\href{https://arxiv.org/abs/1810.02055}{\tt arXiv:1810.02055 [hep-th]}.


\bibitem{Brandao:2017irx}
F.~G.~S.~L.~Brandao, E.~Crosson, M.~B.~Şahinoğlu and J.~Bowen,
``Quantum Error Correcting Codes in Eigenstates of Translation-Invariant Spin Chains,''
\href{https://arxiv.org/abs/1710.04631}{\tt arXiv:1710.04631 [quant-ph]}.


\bibitem{Hayden:2008a}
P.~Hayden, M.~Horodecki, A.~Winter, and J.~Yard,
``A Decoupling Approach to the Quantum Capacity,''
\href{https://doi.org/10.1142/S1230161208000043}{\emph{Open Systems \& Information Dynamics} {\bf 15} (2008) no. 1, 7-19,}
\href{https://arxiv.org/abs/quant-ph/0702005}{\tt arXiv:quant-ph/0702005}.


\bibitem{Horodecki:2008}
M.~Horodecki, S.~Lloyd and A.~Winter,
``Quantum Coding Theorem from Privacy and Distinguishability,''
\href{https://doi.org/10.1142/S1230161208000067}{\emph{Open Systems \& Information Dynamics} {\bf 15} (2008) no. 1, 47-69,}
\href{https://arxiv.org/abs/quant-ph/0702006}{\tt arXiv:quant-ph/0702006}.


\bibitem{Hayden:2008b}
P.~Hayden, P.~Shor and A.~Winter
``Random Quantum Codes from Gaussian Ensembles and an Uncertainty Relation,''
\href{https://doi.org/10.1142/S1230161208000079}{\emph{Open Systems \& Information Dynamics} {\bf 15} (2008) no. 1, 71-89,}
\href{https://arxiv.org/abs/0712.0975}{\tt arXiv:0712.0975 [quant-ph]}.



\bibitem{Hosur:2015ylk}
P.~Hosur, X.~L.~Qi, D.~A.~Roberts and B.~Yoshida,
``Chaos in quantum channels,''
\href{https://doi.org/10.1007/JHEP02(2016)004}{\emph{JHEP} {\bf 1602} (2016) 004},
\href{https://arxiv.org/abs/1511.04021}{\tt arXiv:1511.04021 [hep-th]}.


\bibitem{Hayden:2007cs}
P.~Hayden and J.~Preskill,
``Black holes as mirrors: Quantum information in random subsystems,''
\href{https://doi.org/10.1088/1126-6708/2007/09/120}{\emph{JHEP} {\bf 0709} (2007) 120},
\href{https://arxiv.org/abs/0708.4025}{\tt arXiv:0708.4025 [hep-th]}.


\bibitem{Yoshida:2018ybz}
B.~Yoshida,
``Soft mode and interior operator in Hayden-Preskill thought experiment,''
\href{https://arxiv.org/abs/1812.07353}{\tt arXiv:1812.07353 [hep-th]}.


\bibitem{Yoshida:2019qqw}
B.~Yoshida,
``Firewalls vs. Scrambling,''
\href{https://arxiv.org/abs/1902.09763}{\tt arXiv:1902.09763 [hep-th]}.


\bibitem{DAlessio:2016rwt}
L.~D'Alessio, Y.~Kafri, A.~Polkovnikov and M.~Rigol,
``From quantum chaos and eigenstate thermalization to statistical mechanics and thermodynamics,''
\href{https://doi.org/10.1080/00018732.2016.1198134}{\emph{Adv.\ Phys.}\  {\bf 65} (2016) no.3, 239},
\href{https://arxiv.org/abs/1509.06411}{\tt arXiv:1509.06411 [cond-mat.stat-mech]}.


\bibitem{Deutsch:2018a}
J.~Deutsch,
``Eigenstate thermalization hypothesis,''
\href{https://doi.org/10.1088/1361-6633/aac9f1}{\emph{Rep.\ Prog.\ Phys.}\  {\bf 81} (2018) 082001},
\href{https://arxiv.org/abs/1805.01616}{\tt arXiv:1805.01616 [quant-ph]}.


\bibitem{Garrison:2015lva}
J.~R.~Garrison and T.~Grover,
``Does a single eigenstate encode the full Hamiltonian?,''
\href{https://doi.org/10.1103/PhysRevX.8.021026}{\emph{Phys.\ Rev.\ X} {\bf 8} (2018) no.2,  021026},
\href{https://arxiv.org/abs/1503.00729}{\tt arXiv:1503.00729 [cond-mat.str-el]}.


\bibitem{Dymarsky:2016ntg}
A.~Dymarsky, N.~Lashkari and H.~Liu,
``Subsystem ETH,''
\href{https://doi.org/10.1103/PhysRevE.97.012140}{\emph{Phys.\ Rev.\ E} {\bf 97} (2018) 012140},
\href{https://arxiv.org/abs/1611.08764}{\tt arXiv:1611.08764 [cond-mat.stat-mech]}.


\bibitem{Lashkari:2017hwq}
N.~Lashkari, A.~Dymarsky and H.~Liu,
``Universality of Quantum Information in Chaotic CFTs,''
\href{https://doi.org/10.1007/JHEP03(2018)070}{\emph{JHEP} {\bf 1803} (2018) 070}.
\href{https://arxiv.org/abs/1710.10458}{\tt arXiv:1710.10458 [hep-th]}.


\bibitem{Knill:1996ny}
E.~Knill and R.~Laflamme,
``A Theory of Quantum Error Correcting Codes,''
\href{https://doi.org/10.1103/PhysRevA.55.900}{\emph{Phys.\ Rev.\ A}\  {\bf 55} (2000) 900},
\href{https://arxiv.org/abs/quant-ph/9604034}{\tt arXiv:quant-ph/9604034}.


\bibitem{Beny:2010a}
C.~Bény and O.~Oreshkov,
``General Conditions for Approximate Quantum Error Correction and Near-Optimal Recovery Channels,''
\href{https://doi.org/10.1103/PhysRevLett.104.120501}{\emph{Phys.\ Rev.\ Lett.}\  {\bf 104} (2010) 120501},
\href{https://arxiv.org/abs/0907.5391}{\tt arXiv:0907.5391 [quant-ph]}.



\bibitem{Schumacher:1996a}
B.~Schumacher,
``Sending entanglement through noisy quantum channels,''
\href{https://doi.org/10.1103/PhysRevA.54.2614}{\emph{Phys.\ Rev.\ A}\  {\bf 54} (1996) 2614},
\href{https://arxiv.org/abs/quant-ph/9604023}{\tt arXiv:quant-ph/9604023}.



\bibitem{Beny:2011a}
C.~Bény and O.~Oreshkov,
``Approximate simulation of quantum channels,''
\href{https://doi.org/10.1103/PhysRevA.84.022333}{\emph{Phys.\ Rev.\ A}\  {\bf 84} (2011) 022333},
\href{https://arxiv.org/abs/1103.0649}{\tt arXiv:1103.0649 [quant-ph]}.





\bibitem{Wilde:2015xoa}
M.~Wilde,
``Recoverability in quantum information theory,''
\href{https://doi.org/10.1098/rspa.2015.0338}{\emph{Proc.\ Roy.\ Soc.\ Lond.\ A} {\bf 471} (2015) no.2182,  20150338},
\href{https://arxiv.org/abs/1505.04661}{\tt arXiv:1505.04661 [quant-ph]}.


\bibitem{Junge:2015lmb}
M.~Junge, R.~Renner, D.~Sutter, M.~M.~Wilde and A.~Winter,
``Universal Recovery Maps and Approximate Sufficiency of Quantum Relative Entropy,''
\href{https://doi.org/10.1007/s00023-018-0716-0}{\emph{Annales Henri Poincare} {\bf 19} (2018) no.10,  2955},
\href{https://arxiv.org/abs/1509.07127}{\tt arXiv:1509.07127 [quant-ph]}.


\bibitem{Faist:2019ahr}
P.~Faist, S.~Nezami, V.~V.~Albert, G.~Salton, F.~Pastawski, P.~Hayden and J.~Preskill,
``Continuous symmetries and approximate quantum error correction,''
\href{https://arxiv.org/abs/1902.07714}{\tt arXiv:1902.07714 [quant-ph]}.


\bibitem{Woods:2019fpy}
M.~P.~Woods and Á.~M.~Alhambra,
``Continuous groups of transversal gates for quantum error correcting codes from finite clock reference frames,''
\href{https://arxiv.org/abs/1902.07725}{\tt arXiv:1902.07725 [quant-ph]}.


\bibitem{Hikida:2018khg}
Y.~Hikida, Y.~Kusuki and T.~Takayanagi,
``Eigenstate thermalization hypothesis and modular invariance of two-dimensional conformal field theories,''
\href{https://doi.org/10.1103/PhysRevD.98.026003}{\emph{Phys.\ Rev.\ D} {\bf 98} (2018) no.2,  026003},
\href{https://arxiv.org/abs/1804.09658}{\tt arXiv:1804.09658 [hep-th]}.



\bibitem{Brehm:2018ipf}
E.~M.~Brehm, D.~Das and S.~Datta,
``Probing thermality beyond the diagonal,''
\href{https://doi.org/10.1103/PhysRevD.98.126015}{Phys.\ Rev.\ D {\bf 98} (2018) no.12,  126015},
\href{https://arxiv.org/abs/1804.07924}{\tt arXiv:1804.07924 [hep-th]}.


\bibitem{Datta:2019jeo}
S.~Datta, P.~Kraus and B.~Michel,
``Typicality and thermality in 2d CFT,''
\href{https://arxiv.org/abs/1904.00668}{\tt arXiv:1904.00668 [hep-th]}.



\bibitem{Pastawski:2016qrs}
F.~Pastawski and J.~Preskill,
``Code properties from holographic geometries,''
\href{https://doi.org/10.1103/PhysRevX.7.021022}{\emph{Phys.\ Rev.\ X} {\bf 7} (2017) no.2,  021022},
\href{https://arxiv.org/abs/1612.00017}{\tt arXiv:1612.00017 [quant-ph]}.


\bibitem{Cotler:2017erl}
J.~Cotler, P.~Hayden, G.~Penington, G.~Salton, B.~Swingle and M.~Walter,
``Entanglement Wedge Reconstruction via Universal Recovery Channels,''
\href{https://arxiv.org/abs/1704.05839}{\tt arXiv:1704.05839 [hep-th]}.


\bibitem{Bao:2017guc}
N.~Bao and H.~Ooguri,
``Distinguishability of black hole microstates,''
\href{https://doi.org/10.1103/PhysRevD.96.066017}{\emph{Phys.\ Rev.\ D} {\bf 96} (2017) no.6,  066017},
\href{https://arxiv.org/abs/1705.07943}{\tt arXiv:1705.07943 [hep-th]}.


\bibitem{Choi:2019nhg}
S.~Choi, Y.~Bao, X.~L.~Qi and E.~Altman,
``Quantum error correction and entanglement phase transition in random unitary circuits with projective measurements,''
\href{https://arxiv.org/abs/1903.05124}{\tt arXiv:1903.05124 [quant-ph]}.


\bibitem{Biroli:2010a}
G.~Biroli, C.~Kollath and A.~Läuchli,
``Effect of Rare Fluctuations on the Thermalization of Isolated Quantum Systems,''
\href{https://doi.org/10.1103/PhysRevLett.105.250401}{\emph{Phys.\ Rev.\ Lett.}\  {\bf 105} (2011) 250401},
\href{https://arxiv.org/abs/0907.3731}{\tt arXiv:0907.3731 [cond-mat.quant-gas]}.


\bibitem{Mueller:2015a}
M.~Mueller, E.~Adlam, L.~Masanes and N.~Wiebe,
``Thermalization and canonical typicality in translation-invariant quantum lattice systems,''
\href{https://doi.org/10.1007/s00220-015-2473-y}{\emph{Commun.\ Math\ Phys.}\  {\bf 340} (2011) no.2 499-561},
\href{https://arxiv.org/abs/1312.7420}{\tt arXiv:1312.7420 [quant-ph]}.


\bibitem{Sonner:2017hxc}
J.~Sonner and M.~Vielma,
``Eigenstate thermalization in the Sachdev-Ye-Kitaev model,''
\href{https://doi.org/10.1007/JHEP11(2017)149}{\emph{JHEP} {\bf 1711} (2017) 149},
\href{https://arxiv.org/abs/1707.08013}{\tt arXiv:1707.08013 [hep-th]}.  


\bibitem{Foini:2018sdb}
L.~Foini and J.~Kurchan,
``The Eigenstate Thermalization Hypothesis and Out of Time Order Correlators,''
\href{https://doi.org/10.1103/PhysRevE.99.042139}{\emph{Phys.\ Rev.\ E} {\bf 99} (2019) no.4,  042139},
\href{https://arxiv.org/abs/1803.10658}{\tt arXiv:1803.10658 [cond-mat.stat-mech]}.



\bibitem{Zhuang:2019jyq}
Q.~Zhuang, T.~Schuster, B.~Yoshida and N.~Y.~Yao,
``Scrambling and Complexity in Phase Space,''
\href{https://arxiv.org/abs/1902.04076}{\tt arXiv:1902.04076 [quant-ph]}.



\bibitem{deBoer:2019kyr}
J.~De Boer, R.~Van Breukelen, S.~F.~Lokhande, K.~Papadodimas and E.~Verlinde,
``Probing typical black hole microstates,''
\href{https://arxiv.org/abs/1901.08527}{\tt arXiv:1901.08527 [hep-th]}.


\bibitem{Nie:2018dfe}
L.~Nie, M.~Nozaki, S.~Ryu and M.~T.~Tan,
``Signature of quantum chaos in operator entanglement in 2d CFTs,''
\href{https://arxiv.org/abs/1812.00013}{\tt arXiv:1812.00013 [hep-th]}.


\bibitem{Anous:2019yku}
T.~Anous and J.~Sonner,
``Phases of scrambling in eigenstates,''
\href{https://arxiv.org/abs/1903.03143}{\tt arXiv:1903.03143 [hep-th]}.


\bibitem{Jensen:2019cmr}
K.~Jensen,
``Scrambling in nearly thermalized states at large central charge,''
\href{https://arxiv.org/abs/1906.05852}{\tt arXiv:1906.05852 [hep-th]}.


\bibitem{Kang:2018xqy}
M.~J.~Kang and D.~K.~Kolchmeyer,
``Holographic Relative Entropy in Infinite-dimensional Hilbert Spaces,''
\href{https://arxiv.org/abs/1811.05482}{\tt arXiv:1811.05482 [hep-th]}.


\bibitem{Jahn:2019nmz}
A.~Jahn, M.~Gluza, F.~Pastawski and J.~Eisert,
``Majorana dimers and holographic quantum error-correcting codes,''
\href{https://arxiv.org/abs/1905.03268}{\tt arXiv:1905.03268 [hep-th]}.


\bibitem{Harris:2018jfl}
R.~J.~Harris, N.~A.~McMahon, G.~K.~Brennen and T.~M.~Stace,
``Calderbank-Shor-Steane holographic quantum error-correcting codes,''
\href{https://doi.org/10.1103/PhysRevA.98.052301}{\emph{Phys.\ Rev.\ A} {\bf 98} (2018) no.5,  052301},
\href{https://arxiv.org/abs/1806.06472}{\tt arXiv:1806.06472 [quant-ph]}.


\bibitem{Beny:2009a}
C.~Bény, A.~Kempf and D.~Kribs,
``Quantum error correction on infinite-dimensional Hilbert spaces,''
\href{https://doi.org/10.1063/1.3155783}{\emph{Journal of Mathematical Physics} {\bf 50} (2009) 062108},
\href{https://arxiv.org/abs/0811.0421}{\tt arXiv:0811.0421 [quant-ph]}.


\bibitem{Lidar:2013ed}
D.~Lidar and T.~Brun, editors,
\emph{Quantum Error Correction}
(Cambridge University Press, Cambridge, 2013).









\end{thebibliography}
\end{document}